\documentclass[preprint, 12pt]{elsarticle}
\makeatletter
\def\ps@pprintTitle{%
	\let\@oddhead\@empty
	\let\@evenhead\@empty
	\def\@oddfoot{\centerline{\thepage}}%
	\let\@evenfoot\@oddfoot}
\makeatother
\usepackage{setspace}
\usepackage{times}
\usepackage{siunitx}
\DeclareSIUnit{\wtpercent}{wt.\%}
\usepackage{listings}
\DeclareSIUnit\litre{l}
\usepackage[utf8]{inputenc}
\usepackage{graphicx}
\usepackage{xcolor}
\usepackage{booktabs}
\usepackage{amsmath,amsthm,amssymb,amsfonts,amscd,amsbsy,amsxtra}
\usepackage{wasysym}
\usepackage{ae} 
\usepackage{pgf}
\usepackage{subcaption}
\usepackage{ulem}
\usepackage{lineno}
\usepackage{siunitx}
\usepackage{bbm}
\usepackage{tabularx}
\usepackage{upgreek}
\usepackage{longtable}
\usepackage[longtable]{multirow}
\usepackage{bm}
\usepackage{ragged2e}%rechts- und linksbündig auch mit Umbruch
\newcolumntype{R}[1]{>{\RaggedLeft\arraybackslash}p{#1}}%rechtsbündig
\newcolumntype{L}[1]{>{\RaggedRight\arraybackslash}p{#1}}%linksbündig

\RequirePackage[colorlinks,citecolor=blue,urlcolor=blue]{hyperref}
\usepackage[left=2cm,right=2cm,top=1.5cm,bottom=1.5cm,includeheadfoot]{geometry}

\newcommand{\R}{\mathbb{R}}

\begin{document}
% Comment out, if indenting at the beginning of paragraphs is desired.
\setlength\parindent{0pt}
	
\begin{frontmatter}
%TC:ignore

\title{Quantifying local heterogeneities in the 3D morphology of X-PVMPT battery electrodes based on FIB-SEM measurements}

%\title{Statistical microstructure analysis of an X-PVMPT battery electrode based on 3D FIB-SEM measurements}

%%% https://pubs.acs.org/page/aanmf6/about.html

 \author[1]{Lukas Dodell
 \fnref{eca}
 }
 \author[2]{Matthias Neumann\fnref{eca,cor1}}
 \author[3]{Markus~Osenberg}
 \author[3]{Andr\'{e}~Hilger}
\author[4]
{Gauthier Studer}
 \author[4]
{Birgit~Esser}
 \author[3]{Ingo~Manke}
 \author[1]{Volker~Schmidt}

 \address[1]{Institute of Stochastics, Ulm University, 89069 Ulm, Germany}
\address[2]{
 Institute of Statistics, Graz University of Technology, 8010 Graz, Austria}
 \address[3]{Institute of Applied Materials, Helmholtz Center for Materials and Energy, 14109 Berlin, Germany}
 \address[4]{Institute of Organic Chemistry II and Advanced Materials, Ulm University, 89069 Ulm, Germany}

 \fntext[eca]{LD and MN  contributed equally to this paper.}
\fntext[cor1]{Corresponding author. Email: neumann@tugraz.at. Phone: +43 316 873 - 4543}

%Maximum of 200 words
\begin{abstract}
Organic electrode-active materials (OAMs) not only enable a variety of charge and storage mechanisms, but are also safer for the environment and of lower cost compared to materials in commonly used lithium-ion batteries. Cross-linked Poly(3)-vinyl-\textit{N}-methylphenothiazine (X-PVMPT) is a p-type OAM which shows high performance and enables fast and reversible energy storage in different battery configurations. The performance of an OAM does not only depend on its molecular or polymer structure, but also on the structure of the composite electrode. The porous nanostructure of an electrode composed of X-PVMPT, a conductive carbon additive and binder is investigated by statistical image analysis, based on 3D image data obtained by focused-ion beam scanning-electron microscopy (FIB-SEM) measurements. Univariate probability distributions of relevant morphological descriptors as well as bivariate distributions of pairs of such descriptors are parametrically modelled, among others, by utilization of copulas in the latter case. These models 
are then used for quantifying local heterogeneities of X-PVMPT  considered in this paper. 
%showed equivalent behaviour as the corresponding empirical distributions estimated from the data. 
Furthermore, it is shown that the nanostructure changes when traversing from bottom to top face of the electrode, which influences its performance. While the observed short transportation paths trough the solid phase are beneficial in terms of electrical conductivity, the pathways through the pore phase influencing the effective ionic diffusivity are--in comparison--rather long.

\end{abstract}

% Maximum of 6 Keywords
%TC:endignore
\end{frontmatter}

%\linenumbers
\modulolinenumbers[5]

\section{Introduction}\label{sec:introduction}
Electrochemical energy storage is of increasing relevance in today’s technology-driven world with an ever-rising global demand for batteries. This includes electromobility and consumer electronics, but also stationary grid-storage, where low-cost and safer alternatives to the lithium-ion battery might be future technologies. Organic electrode-active materials (OAMs) are attractive candidates for alternative battery concepts, made from more abundant elements~\cite{kimJ1}, being potentially cheaper and safer, and enabling a variety of charge (and concurrent ion) storage mechanisms~\cite{esser2,esser3,Poizot4}. Functioning under uptake of electrolyte (metal) cations (for n-type OAMs) or anions (p-type OAMs), OAMs can be used as a replacement for metal oxides in classical metal-ion batteries \cite{yang5}, but also function well with multivalent metal ions~\cite{huang6,li7,chen8} as well as in anionic cell configurations~\cite{poizot9}. Poly(3-vinyl-\textit{N}-methylphenothiazine) (PVMPT) or its cross-linked form X-PVMPT are high-performance p-type OAMs that enable fast and reversible energy storage (under anion insertion) in batteries at high potential of 3.5 V vs Li/Li+~\cite{Kolek10}. The charge storage mechanism~\cite{kolek11,otteny12}, effect of electrolyte type~\cite{perner13} and conductive carbon additive were studied in detail for PVMPT~\cite{tengen14}. Due to its diminished solubility in liquid battery electrolytes~\cite{otteny15}, cross-linked X-PVMPT is a particularly relevant OAM for full cells and showed exceptional performance in Al-based batteries~\cite{studer16}, surpassing graphite in specific capacity with high rate performance, as well as in an all-organic anion-rocking chair battery~\cite{bhosale17}. Structural engineering by changing the polymer backbone~\cite{desmaizieres18} effects solubility~\cite{otteny19} and charge storage mechanism as well as rate capability~\cite{otteny19,acker20}, processing abilities~\cite{acker21} and applicability~\cite{wessling22,Delgado23,wessling.2024} of phenothiazine-based redox polymers~\cite{otteny24}. 

Decisive for the performance of an OAM is not only its molecular or polymer structure, but to a large extent the morphology of the composite electrode used in the battery. The composite electrode consists of the OAM as well as a conductive carbon additive and binder. Its morphology strongly influences electron- as well as ion transport, effecting the charge-storage performance. We herein investigate an X-PVMPT-based composite electrode with respect to the 3D morphology of its nanopores. For this purpose, 3D imaging is combined with methods from spatial statistics, mathematical morphology and machine learning in order to quantify the nanostructure of the considered X-PVMPT-based composite electrode. This allows us to locally evaluate porosity or surface area per unit volume as well as to determine morphological descriptors, which are not accessible experimentally. For polymer-based battery electrodes, this approach has been used to elucidate morphological differences between PTMA, i.e.  poly(2,2,6,6-tetramethyl-4-piperinidyl-$N$-oxyl methacrylate)-based, electrodes manufactured either with polyvinylidenfluoride (PVDF) or with water-soluble binder~\cite{Neumann23}. Recently, this methodology also revealed the impact of cross-linking in PTMA-CMK8 electrodes on their morphology~\cite{Ademmer2023}. In both studies, the morphology on the micrometre-scale was quantified based on 3D-image data acquired by synchrotron-tomography~\cite{heenan.2019, tang.2021}. 

In the present paper, we consider the 3D morphology of an X-PVMPT electrode on the nanometer-scale. Thus, focused-ion beam scanning-electron microscopy (FIB-SEM)~\cite{holzer.2012, moebus.2007} is used for image acquisition which allows for a voxel resolution of $10~\mathrm{nm}.$ As in~\cite{Ademmer2023}, image segmentation, i.e. the detection of pore voxels based on greyscale images, is performed by machine learning using the software ilastik~\cite{berg.2019}. The focus of the subsequent statistical image analysis is on the quantification of local heterogeneities. Going beyond the descriptive analysis performed in~\cite{Neumann23, Ademmer2023}, we use parametric statistics to model both the univariate distribution of local morphological descriptors as well as bivariate distributions of pairs of such descriptors. Doing so, the heterogeneity of the complex nanoporous morphology is characterized by a small number of model parameters. Moreover, by parametric modeling of the bivariate distributions, we obtain analytical formulas for the corresponding conditional distributions, such as, e.g., the distribution of the local surface area per unit volume conditioned on a predefined value of  local porosity. From the conditional distributions, in turn, one can easily derive the corresponding expectations, variances or quantiles, which elucidate quantitative relationships between pairs of local morphological descriptors. For this modeling approach we build upon methodologies previously developed for quantifying the 3D morphology of paper-based  materials~\cite{n.2020, neumann.2024}. Furthermore, structural gradients within the X-PVMPT-based composite electrode considered in the present paper are revealed.   

%The rest of this paper is organized as follows. In  Section~\ref{sec:matPrepAndImaging} a description of  materials and imaging techniques is given, whereas Section~\ref{sec:imagePreprocessing} deals with pre-processing and segmentation of image data. Section~\ref{sec:four} briefly explains the mathematical methods used for the analysis and modeling of segmented image data. The obtained results are presented in Section~\ref{sec:microstructureAnalysis}. Section~\ref{sec:conclusion} concludes the paper.

\section{Materials and 3D Imaging}\label{sec:matPrepAndImaging}

\subsection{Description of electrode material}\label{subSec:preparation}

The crosslinked X-PVMPT polymer with $\SI{10}{mol}\%$ of the crosslinker was synthesized using the same procedure as reported in \cite{otteny15,studer16}. Composite electrodes were prepared using $\SI{50}{\wtpercent}$ X-PVMPT, $\SI{45}{\wtpercent}$ carbon additive (Super C65, Timcal) and $\SI{5}{\wtpercent}$ PVdF (Kynar HSV 900, Arkema). X-PVMPT and carbon black were pre-dried in a vacuum oven ($\SI{e-3}{\milli\bar}$) at $\SI{60}{\celsius}$ for $\SI{24}{\hour}$, then pre-mixed using a planetary centrifugal mixer (1500 rpm, $\SI{15}{\minute}$, ARM 310, Thinky mixer). The binder was added as a $\SI{5}{\wtpercent}$ PVDF in $N$-methyl-2-pyrrolidone (NMP, Acroseals, Thermo scientific, $99.5\%$, stored over molecular sieves) solution, and the composite mixed. Finally, stepwise NMP additions and mixing steps (1500 rpm, $\SI{15}{\minute}$) were performed until a honey-like viscosity was obtained. The resulting paste was blade-coated onto KOH-etched aluminum foil ($1235$ aluminum foil, H18 hard state, $\SI{20}{\micro\meter}$ from Gelon LIB), with a wet-film thickness of $\SI{100}{\micro \meter}$.  The coated foil was dried at ambient pressure at $\SI{60}{\celsius}$ for $\SI{24}{\hour}$, and electrodes with a diameter of $\SI{12}{\milli \meter}$ were punched out with an electrode-punching device (EL-Cut from EL-CELL). The active material (X-PVMPT) mass loadings of the electrodes lay between $\SI{0.22}{\milli\gram /\centi\meter^{-2}}$ and $\SI{0.24}{\milli\gram /\centi\meter^{-2}}$. Note that the theoretical specific capacity of X-PVMPT for a one-electron process is $112~\mathrm{mAh/g}$, while the practical capacity lies close to this value in Li-half cells~\cite{otteny15}.

\subsection{3D imaging by FIB-SEM tomography}\label{subSec:synchrotronImaging}
For 3D tomography using FIB-SEM, the polymer sample was first mounted on a standard $12.5~\mathrm{mm}$ aluminium SEM stub. For this purpose, a $1~\mathrm{mm} \times 3~\mathrm{mm}$ piece was cut out with a scalpel and glued to the SEM holder with the aluminum conductor side using a carbon adhesive pad. A copper tape was glued over one side of the sample to ensure good conductivity of the sample during SEM imaging.
The sample was then inserted into the ZEISS Crossbeam 340 of the Corelab Correlative Microscopy and Spectroscopy (CCMS) at Helmholtz-Zentrum Berlin (HZB). Using a gallium ion beam current of $15~\mathrm{nA}$ (at $30~\mathrm{keV}$), the polymer material and parts of the underlying aluminum conductor were first removed in a $30~\upmu \mathrm{m} \times 30~\upmu \mathrm{m}$ area in the centre of the polymer sample. The cross-section exposed for imaging was polished with an ion current of $300~\mathrm{pA}$. A low acceleration voltage of $1~\mathrm{keV}$ was used for imaging to minimize the depth of electron penetration into the sample. An image size of $2048~ \mathrm{pixels} \times 1536~\mathrm{pixels}$ was selected for imaging with an acquisition time of $20.5$  seconds per image. A chamber mounted SE2 detector was used for serial imaging. With a field of view of $20.5~\upmu \mathrm{m}$, this resulted in a pixel size of $10~\mathrm{nm}$. The ion beam was then used to remove a $10~\mathrm{nm}$ layer of the sample between each image using the $300~\mathrm{pA}$ current at $30~\mathrm{keV}$, resulting in a 3D data set with an isotropic voxel edge length of $10~\mathrm{nm}$.

\section{Pre-processing and segmentation of image data}\label{sec:imagePreprocessing}

\begin{figure}[h]
\center
\includegraphics[width=0.6\textwidth]{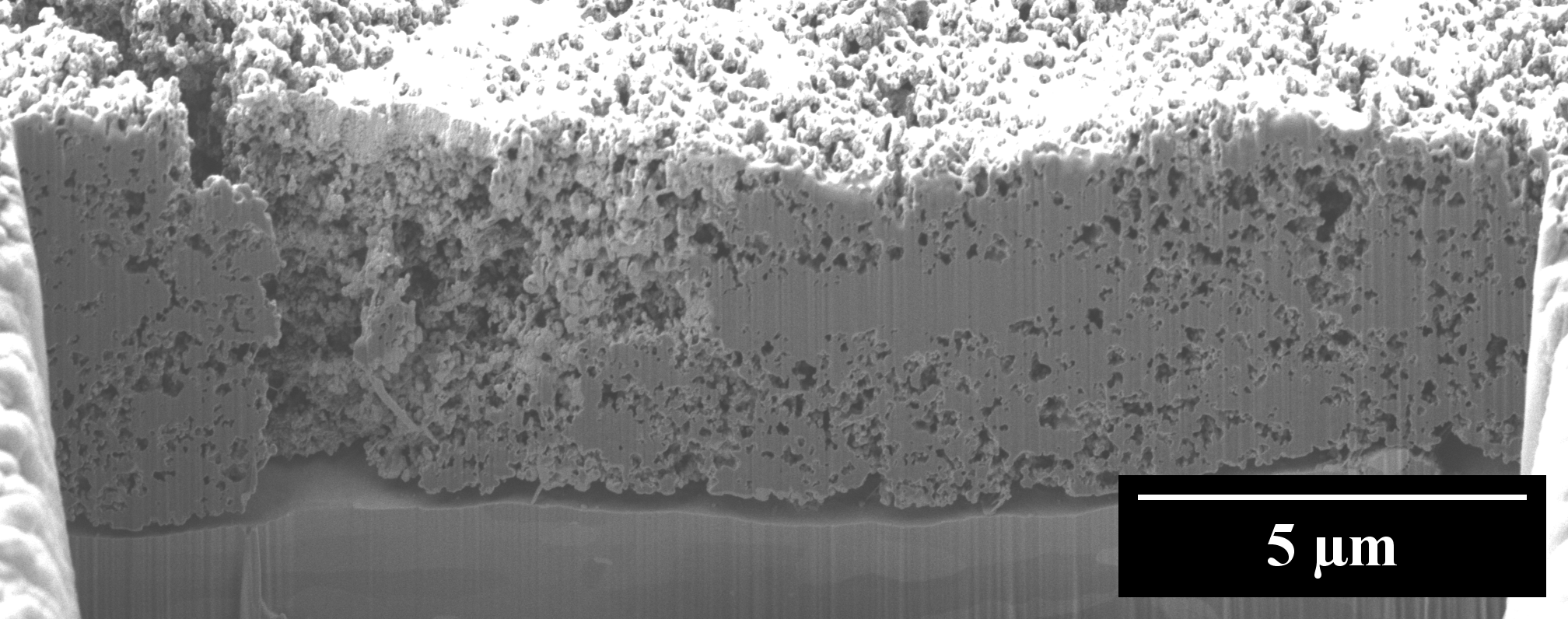}\\
\caption{2D Slice of greyscale image data obtained by FIB SEM tomography.}
\label{fig:slice}
\end{figure}

Using the results of the FIB-SEM measurements  described in Section~\ref{subSec:synchrotronImaging}, the porous nanostructure of  the X-PVMPT-based  electrode material was reconstructed from 3D-greyscale image data by a phase-based segmentation. Prior to the segmentation, pre-processing of the greyscale image data was required. First, a drift correction was applied. This was done by registering the areas above the current cross-section. These areas had not yet been cut by the gallium beam and therefore contained image information about the surface of the unaltered sample. Therefore, features detected in these areas had to be constantly moving towards the cutting edge of the forward moving cross-section. Any non-constant motion was corrected using the scale-invariant feature transform (SIFT) algorithm~\cite{lowe.2004}, implemented in FIJI~\cite{Fiji2012}. Note that this correction has been applied to the greyscale image data. Subsequently, the
sample is aligned to the coordinate system of the sampling window, as shown in Figure~\ref{fig:slice}. The $x$-axis goes from left to right, the $z$-axis from bottom to top, and the $y$-axis goes through the image
slices starting at the first image slice. \\

%TC:ignore
%TC:endignore

\begin{figure}[ht]
\center
        \begin{subfigure}[c]{0.32\textwidth}
		\center
		\includegraphics[width=1\textwidth]{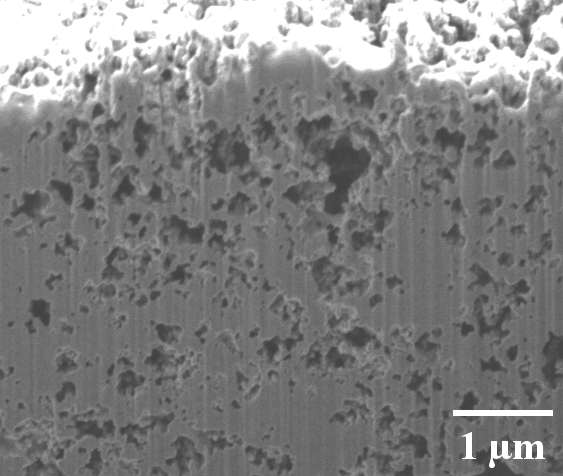}\\
		(a)
	\end{subfigure}
        \begin{subfigure}[c]{0.32\textwidth}
		\center
		\includegraphics[width=1\textwidth]{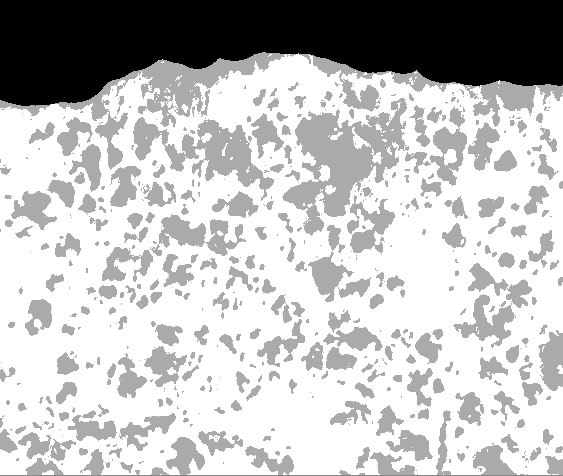}\\
		(b)
	\end{subfigure}

	\begin{subfigure}[c]{0.32\textwidth}
		\center
		\includegraphics[width=1\textwidth]{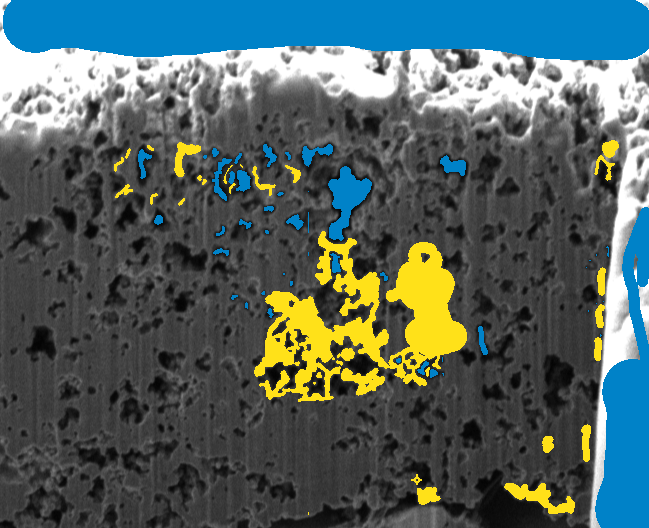}\\
		(c)
	\end{subfigure}
    \begin{subfigure}[c]{0.22\textwidth}
		\center
		\includegraphics[width=1\textwidth]{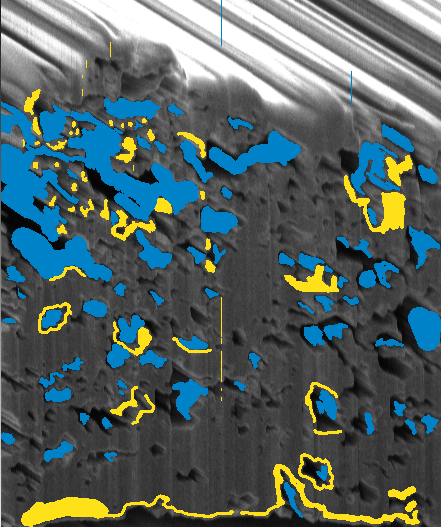}\\
		(d)
	\end{subfigure}
    \begin{subfigure}[c]{0.30\textwidth}
		\center
		\includegraphics[width=1\textwidth]{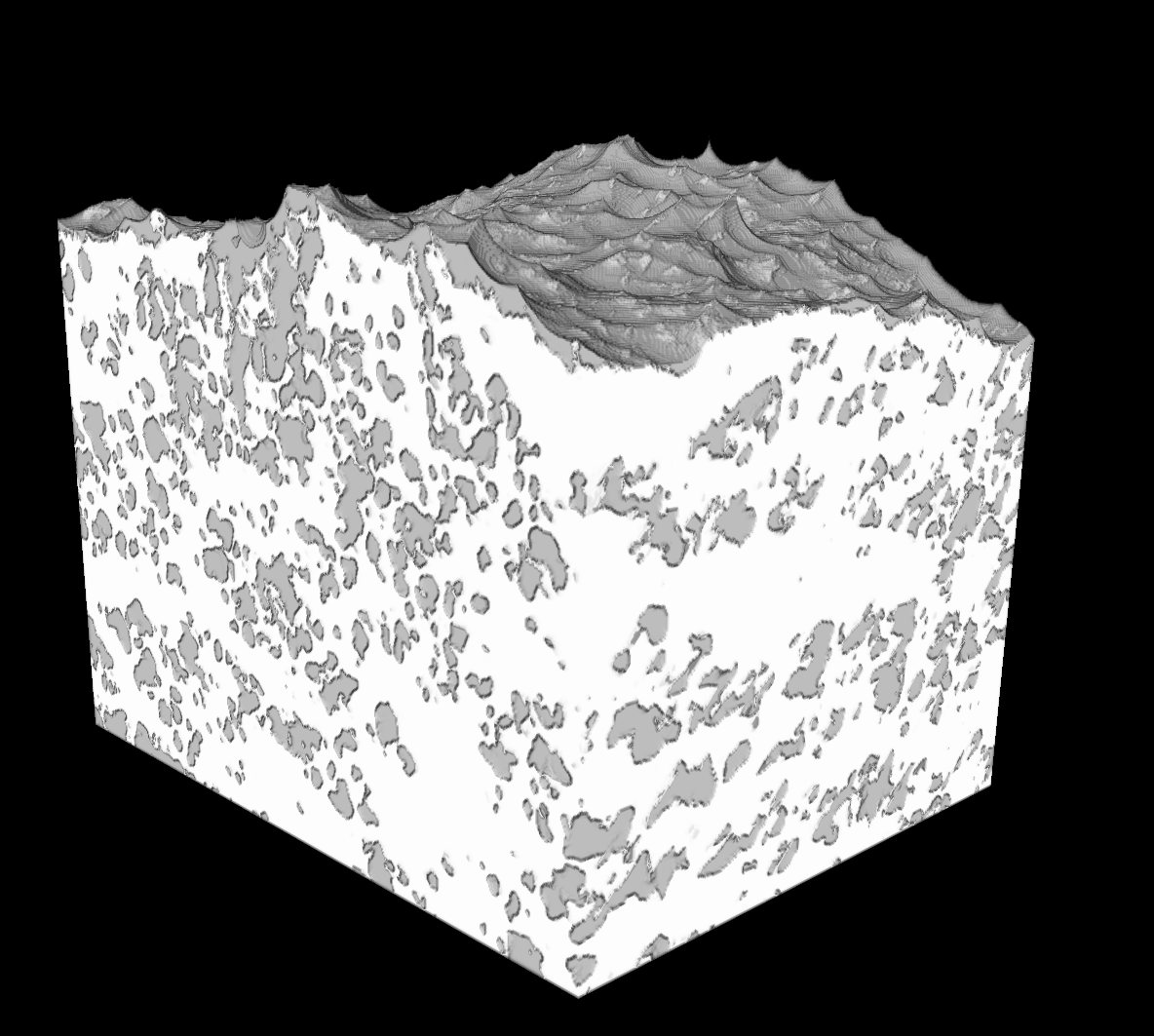}\\
		(e)
	\end{subfigure}
	\caption{Cutout of FIB-SEM image data (a), corresponding 2D slice of segmentation into solid-phase (white), pores (grey) and background (black) (b), hand-labeling of  image data: blue  color represents background and pores, whereas yellow represents the solid phase (c,d), 3D rendering of the segmented image data (e)} \label{fig:labelling}
\end{figure}

In the next step, to analyze the nanostructure of the electrode, a phase-based segmentation of the underlying greyscale image data obtained by FIB-SEM tomography was performed. This means that each voxel of the image is assigned to either solid, pores or background. Note, however,  that the electrode material depicted in the image data has a large diagonal crack, 
 see  Figure~\ref{fig:slice}. 
This is not only problematic in terms of segmentation, but also for the subsequent statistical analysis of the segmented nanostructure. Therefore, only a cutout, visualized in Figure~\ref{fig:labelling}a, was used for statistical image analysis. To segment the cutout into solid phase and the union of pores and background, a random forest classifier~\cite{breiman.2001} was trained using the software ilastik \cite{berg.2019}. The training was performed based on hand-labeled data. Two hand-labeled slices are shown in Figures~\ref{fig:labelling}c and \ref{fig:labelling}d, where in the latter case a 2D slice is depicted which is orthogonal to that shown in Figure~\ref{fig:labelling}c. The trained random forest allows us to decide for each voxel whether it belongs to the solid phase. Details are provided in the supplementary information.

Finally, to distinguish between background and pores in the complement of the solid phase, a so-called rolling ball algorithm \cite{machado} with a radius of 20 voxels ($0.2~\upmu\mathrm{m}$) has been used, which was previously  applied for PTMA-based battery electrodes~\cite{Neumann23,Ademmer2023}. A 2D slice of the final segmentation is shown in Figure~\ref{fig:labelling}b. \\ %Based on this segmentation, the statistical analysis for characterizing the morphology of the nanostructure is performed in Section~\ref{sec:microstructureAnalysis}. 

% \begin{figure}[ht]
% \center
% \includegraphics[width=0.32\textwidth]{images/Substackcutout (220).png}\\
% \caption{Cutout of the image data}
% \label{fig:cutout}
% \end{figure}

%TC:ignore
%TC:endignore

% \begin{figure}[ht]
%     \centering
%     \includegraphics[width=0.32\linewidth]{images/Substack (214).png}
%     \caption{Example slice of the segmentation into solid-phase,pore-phase and background}
%     \label{fig:segmentation}
% \end{figure}

\section{Statistical analysis and modeling of segmented image data}\label{sec:four}

\subsection{Morphological descriptors}\label{sec:morphological_descriptors}

 Using tools of statistical image analysis, the segmentation of image data into solid  and pore phases allows for a quantitative characterization of  electrode nanostructures and, in particular, local heterogeneities thereof.  For this purpose,  morphological descriptors of the nanostructure are globally and locally evaluated. The morphological descriptors considered in the present paper are the thickness $\delta$, the porosity $\varepsilon$, the surface area between solid and pore  phases per unit volume $S$ as well as the mean geodesic tortuosity of solid- and pore-phase, denoted by $\tau_s$ and $\tau_p$, respectively. These descriptors are crucial for battery electrodes, since thickness influences the capacity of the electrode, electrochemical reactions take place at the surface area, while porosity and mean geodesic tortuosities strongly influence effective transport properties such as effective ionic diffusivity and effective electric conductivity. Quantitative relationships of the latter are discussed in Chapter~5 of~\cite{holzer.2023}.

 Global morphological  descriptors are computed based on the complete image as sampling window. For the computation of local descriptors we consider non-overlapping square-shaped cutouts with side length of $1~\upmu\mathrm{m}$, the centers of which are arranged on a regular hexahedral grid, in the plane parallel to the aluminum foil. For each of these cutouts the above mentioned morphological descriptors are determined, where the full thickness is taken into account. Doing so, we follow the previously used approach for investigating local heterogeneities in polymer-based batteries~\cite{Neumann23, Ademmer2023} and paper-based materials~\cite{n.2020,neumann.2024}. Then, we get   empirical distributions of local morphological descriptors, which are parametrically modeled as stated in Section~\ref{sec:model} below.

For a given sampling window, the thickness of the electrode $\delta$ is computed as follows. For each line of voxels, which is orthogonal to the aluminum foil, the corresponding thickness is defined as the maximum distance between two voxels on this line belonging to the solid phase of the electrode. The average of these distances over all lines contained in the sampling window is considered to be the thickness of the electrode in this sampling window. The porosity $\varepsilon$ is determined by the point-count method \cite{chiu2013stochastic}, i.e., the porosity is the ratio of pore voxels over all voxels of the electrode within the sampling window. The surface area per unit volume $S$ is computed as the surface area between solid and pores in the considered sampling window divided by the volume of the sampling window. For the computation of surface areas from voxelized image data, we use the algorithm proposed in~\cite{schladitz}. Finally, we consider the mean geodesic tortuosity which is a purely geometrical descriptor quantifying the windedness of shortest transportation paths through a given phase. It is defined as the expected shortest path length from a predefined starting point at the bottom of the electrode to the top of the electrode through the phase of interest, which is normalized by the local thickness of the electrode at the starting point. Note that there are several types of tortuosity investigated in the literature~\cite{holzer.2023, clennell.1997, ghanbarian.2013}. Even if the definition of mean geodesic tortuosity considered in the present paper is a purely geometrical descriptor of electrode nanostructure, it strongly correlates with effective transport properties such as effective diffusivity, effective conductivity or permeability~\cite{neumann.2020, prifling.2021b}. To compute the mean geodesic tortuosity of a phase in a given sampling window, the shortest pathways through this phase are determined by applying the Dijkstra algorithm \cite{thulasiraman1992graphs} on the voxel grid. When computing mean geodesic tortuosity on local cutouts as sampling windows, the starting points of the paths are located  within these local cutouts, while the paths themselves are allowed to leave the sampling window in order to avoid a strong influence of edge effects.

%To compute the microstructure descriptors locally, the image data was separated into uniform cuboids referred to as sampling windows in the following way. A hexahedral grid of points was superimposed on the $xz$-plane of the image data, where each point has a distance of 51 Voxels (or $0.51\si{ \mu m}$) to any neighbouring point. These gridpoints are the centers of non-overlapping sampling windows, which extend over the whole thickness of the sample in $y$-direction. This led to 71 sampling windows on which the local microstructure descriptors where then computed. Note that a local descriptor of a sampling window is the average of all descriptor results in the respective sampling window. The mean computed across all sampling windows is then equal to the mean of the whole sample, which we consider as global descriptors.
%Other global descriptors such as the porosity dependent on the $y$-coordinate as well as the boundary length of the solid phase dependent on the $y$-coordinate are computed by appyling the point count method to each $xz$-plane in the first case and summing up the boundary lengths of the solid phase in the $xz$-plane in the latter case. 
%From the results of computed microstructure descriptors we were able to perform a parametric modeling of the descriptors respective probability distribution, which is described in the following section.

\subsection{Parametric modeling of local morphological descriptors}\label{sec:model}

\subsubsection{Univariate  probability densities of single descriptors}\label{sec:uni.den.sin}
For fitting parametric probability distributions to  empirical distributions of the computed local descriptors, namely porosity $\varepsilon$, thickness $\delta$, surface area per unit volume $S$ and mean geodesic tortuosity of solid-phase and pore-phase $\tau_s$ and $\tau_p$, respectively, we proceed as follows. For each descriptor, we consider various families of parametric distributions as candidates. For each of these families, parameter fitting is performed by maximum likelihood estimation. Finally, we choose the distribution where the maximum likelihood is largest  over all candidate families of distributions. At this, the maximum likelihood is penalized by the numer of fitted parameters by means of Akaike's information criterion \cite{hamparsum.1987}. In this way,   for the five  local morphological descriptors mentioned above, we selected a
 Maxwell-Boltzmann distribution $\mathsf{MB}(\mu_{M}, \sigma_{M})$,  Rayleigh distribution $\mathsf{R}(\mu_R, \sigma_R)$ and shifted Gammma distribution $\Gamma(a_\Gamma, \sigma_\Gamma, \mu_\Gamma)$, as well as mixtures of Gaussian  and beta  distributions, denoted by $\mathsf{N}(\mu_1, \mu_2, \sigma_1, \sigma_2, p_N)$ and $\mathsf{Beta}(\alpha_1, \alpha_2, \beta_1, \beta_2, p_B)$ for parameters $\mu_{M}, \mu_R,  \mu_\Gamma, \mu_1, \mu_2 \in \R, \sigma_{M}, \sigma_R, a_\Gamma, \sigma_\Gamma, \sigma_1, \sigma_2, \alpha_1, \alpha_2, \beta_1, \beta_2 > 0$ and $0 \leq p_N, p_B \leq 1.$ The probability density functions of these distributions are provided in the supplementary information. The fitted parameters are presented in Table~\ref{tab:fam} below.

%\subsubsection{Measures of local heterogeneity; Correlation matrix}
 %To analyze the microstructure of the electrode we want to look at measures that further quantify the local heterogeneity. Therefore we look at the correlation matrix in Table~\ref{tab:correlations-matrix}, which measures the linear dependence of two variables. The resulting correlation coefficients for porosity and every other microstructure characteristic indicate that these characteristics are linearly dependent on each other. For this reason, the modelling of bivariate distributions of descriptor pairs is sensible. 

%\begin{table}[ht]
%\begin{center}
%\begin{tabular}{cccccc}
 %    & volfrac & thickness & specsurf & tortpore & tortsolid \\ \hline
 %CV    & 0.1813 & 0.0573 & 0.1364 & 0.0507 & 0.0156 
%\end{tabular}
%\caption{Coefficient of variation for each microstructure descriptor}
%\label{tab:cov}

%\end{center}
%\end{table}

\subsubsection{Copula-based bivariate densities of descriptor pairs}\label{sec:biv.dis}

The bivariate probability densities of descriptor pairs
could simply be modelled as products of the corresponding univariate densities considered in Section~\ref{sec:uni.den.sin}. However, this would require that the underlying local morphological descriptors would not be  (or, at least be only weakly) correlated. But, this is not the case, as can be seen in 
Table~\ref{tab:correlations-matrix}.

\begin{table}[h]
\begin{center}
\begin{tabular}{cccccc}
  & $\varepsilon$ & $\delta$ & $S$ & $\tau_s$ & $\tau_p$\\ \hline
  $\varepsilon$ & 1.0 & 0.55 & 0.86 & -0.59&  0.80 \\ \hline
  $\delta$ &  & 1.0 & 0.58 & -0.48 & 0.46\\ \hline
  $S$ &  &   & 1.0 & -0.64 & 0.59 \\ \hline
  $\tau_s$ &  &  &   & 1.0 & -0.46\\ \hline
  $\tau_p$ & & & & & 1.0   
\end{tabular}
\caption{Correlation matrix of local morphological descriptors $\varepsilon$, $\delta$, $S$, $\tau_s$ and $\tau_p$.}
\label{tab:correlations-matrix}
\end{center}
\end{table}

Moreover, in most cases, the 
univariate densities considered in Section~\ref{sec:uni.den.sin} are not Gaussian, 
 see Figures~\ref{fig:univ1} and~\ref{fig:univ2}. Therefore, the bivariate probability densities of descriptor pairs are modelled   by means of so-called copulas \cite{n.2020}.
 This approach is advantageous in a sense that the complexity of the model is split into modeling the univariate marginal distributions and modeling the copula, where the copula  contains the information on the interdependence of the individual morphological descriptors.  % Moreover, the copula approach is rather flexible as it allows to model bivariate distributions with given univariate marginal distributions.

For modeling the  bivariate probability densities of despriptor pairs consisting of  local porosity and one of the remaining four  morphological descriptors, we consider one-parametric Archimedean copulas~\cite{nelsen.2007} as model type. It  turned out that in all four cases the best fit was obtained by a Frank copula $C:[0,1]^2\rightarrow[0,1]$, defined by 
$$ 
C\left(u,v\right) = -\frac{1}{\theta}\log\left(1 + \frac{\left(e^{-\theta u}-1\right)\left(e^{-\theta v}-1\right)}{e^{-\theta}-1}\right)
$$
for any $u,v\in[0,1]$, where  $\theta \in \mathbb{R}\setminus \{0\}$ is some model parameter. Plugging the parametric distribution functions and probability densities of the univariate distributions as well as the parametric copula into the formulas for conditional probability densities, we obtain analytical formulas for the latter ones. These analytical formulas fully quantify the relationship between pairs of local morphological descriptors. In particular, they can be used to compute conditional expectations and quantiles of a morphological descriptor for fixed porosity. The latter can be used for predictions of local morphological descriptors based on local porosity.

%The Clayton copula is defined by
%$$ C_{Cla yton}\left(u,v\right) = \left(u^{-\theta} + v^{-\theta} -1\right)^{-\frac{1}%{\theta}} \quad \text{with } \theta \in [-1,\infty)\setminus{\{0\}}  $$
%for all $0 \leq u,v\leq 1$.

\section{Results and discussion}\label{sec:microstructureAnalysis}

\subsection{Results}\label{sec:results}

\begin{figure}[h]
\center
	\begin{subfigure}[c]{0.32\textwidth}
		\center
		\includegraphics[width=1\textwidth]{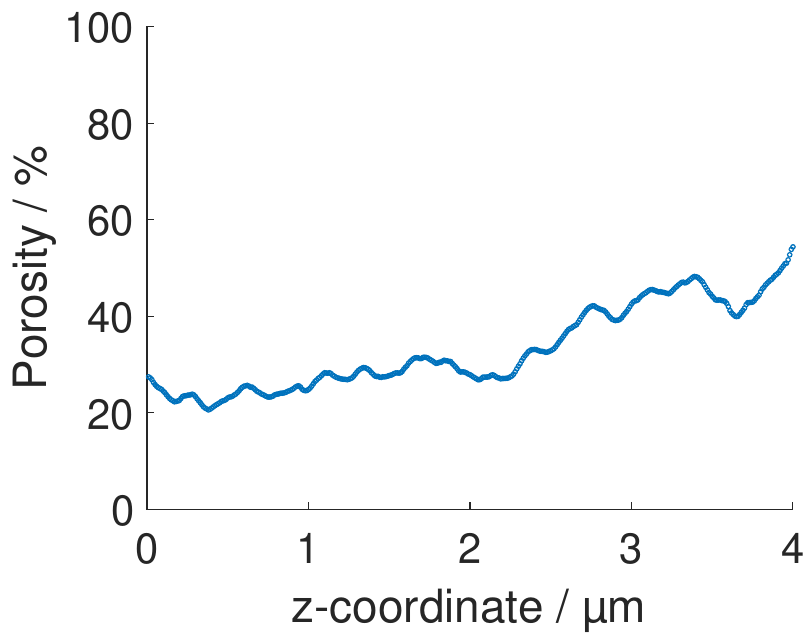}\\
		(a)
	\end{subfigure}
	\begin{subfigure}[c]{0.32\textwidth}
		\center
		\includegraphics[width=1\textwidth]{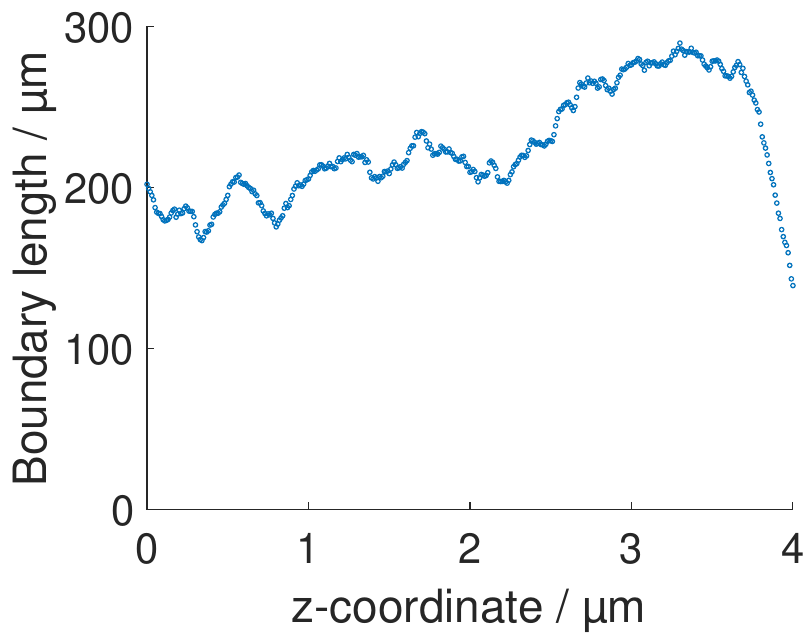}\\
		(b)
	\end{subfigure}
       
	\caption{Porosity (a) and boundary length (b) in dependence on the height $z$ within the electrode. 
%\textcolor{red}{\textbf{MN: @Lukas: In Figure 3 we should write $z$-coordinate on the x-axis. Please replace voxel units by physical units.}}
}  \label{fig:globalY}
\end{figure}

From visual inspection of the cutout shown in Figure \ref{fig:labelling}a it seems that the porosity of the sample  depends on the height, i.e., the $z$-value of the location we look at. This is  confirmed by the results of statistical image analysis, when we determine porosity and boundary length of the interface between pores and solid at a given height $z\in[0,4 ]\upmu \mathrm{m}$, see Figure~\ref{fig:globalY}. Note that for this purpose, we consider the complete sample in the $xy$-plane.  Here we can see that both characteristics strongly  depend  on the $z$-position as porosity and boundary length increase approaching the top part of the electrode.

%TC:ignore
%TC:endignore

Furthermore, the spatial gradient of the boundary length presented in Figure~\ref{fig:globalY} shows that the surface area per unit volume in 3D is increasing with an increasing distance to the aluminum foil. This observation is in good accordance with the spatial gradient observed for porosity as--depending on the shape of the phases--the surface area typically takes its maximum when both phases, pores and solid, have nearly the same volume fraction. For the PTMA-based electrodes considered in~\cite{Neumann23}, e.g., the surface area per unit volume on the micro-scale takes its maximum at a volume fraction of PTMA at $60\%$. 
The values of global morphological descriptors
for the X-PVMPT-based composite electrode considered in the present paper, which have been obtained
by averaging over results for all sampling windows, are given in Table~\ref{tab:Global}.

% Considering the global microstructure descriptors summarized in Table \ref{tab:Global} alongside  \\

\begin{table}[h]
\begin{center}

\begin{tabular}[h]{c|c|c|c|c}
%{L{3cm}|L{3cm}|L{3cm}|L{3cm}|L{3cm}}
$\varepsilon$  & $\delta$   & $S$  & $\tau_s$ & $\tau_p$ \\
\hline
 37 \% &  4.02 \si{\upmu m}  &  8.82 \si{\upmu m^{-1}} &  1.04 &  1.24 \\

\end{tabular}

\caption{Global morphological descriptors, obtained by averaging over results for all sampling windows.}
\label{tab:Global}
\end{center}
\end{table}

\subsubsection{Univariate distributions of local morphological descriptors}\label{sec:univdist}
To investigate  local heterogeneities of  the X-PVMPT-based composite electrode we first computed 
the values of $\varepsilon$, $\delta$, $S$, $\tau_s$
 and $\tau_p$ for each of the
non-overlapping square-shaped cutouts with side
length of 1 µm. The corresponding empirical distributions
were obtained by kernel density estimation using a Gaussian kernel. The  fitted parametric families of distributions and the corresponding parameter values are summarized in Table~\ref{tab:fam}. Note that in the following figures the empirical distributions are referred to as ``data'', the fitted parametric univariate distributions as ``model''.

\begin{table}[h]
\begin{center}

\begin{tabular}[h]{ccc}
descriptor & distribution type & parameter values \\ \hline
$\varepsilon$ &  Beta mixture & $\alpha_1= 109.43$, $\alpha_2= 125.79$, $\beta_1= 65.44$, $\beta_2= 47.74$, $p_B=0.59$ \\ \hline
$\delta$ &  Maxwell-Boltzmann & $\mu_M= 3.46$, $\sigma_M= 0.35$\\ \hline
$S$ & Gaussian mixture & $\mu_1= 13.1$, $\mu_2= 10.46$, $\sigma_1= 0.87$, $\sigma_2= 1.06$, $p_N=0.56$ \\ \hline
$\tau_s$ &  Rayleigh & $\mu_R= 1.01$, $\sigma_R= 0.024$ \\ \hline
$\tau_p$ & Gamma & $a_\Gamma= 2.97$ $\mu_\Gamma= 1.14$, $\sigma_\Gamma=0.037$  \\

\end{tabular}

\caption{Morphological descriptors, fitted  parametric distribution types, and parameter values}
\label{tab:fam}
\end{center}
\end{table}

\begin{figure}[ht]
\center
	\begin{subfigure}[c]{0.32\textwidth}
		\center
		\includegraphics[width=1\textwidth]{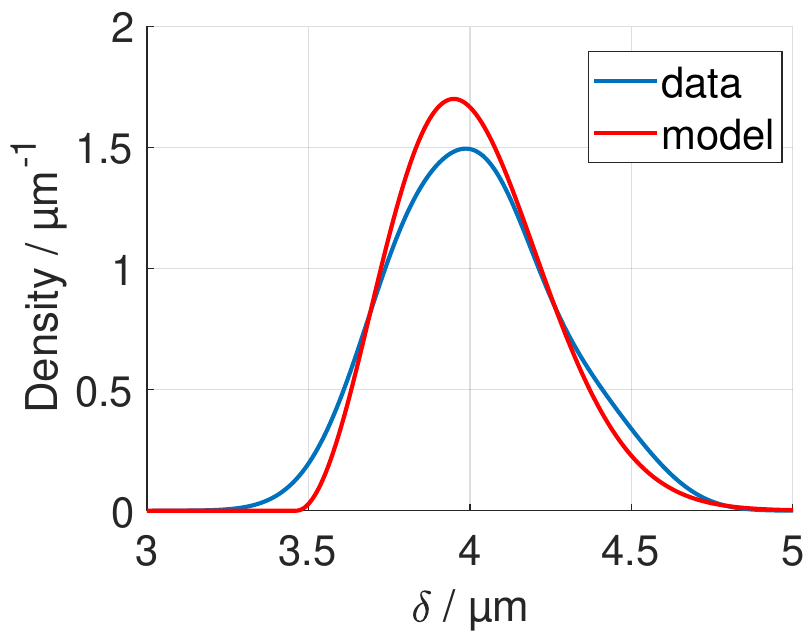}\\
		(a)
	\end{subfigure}
	\begin{subfigure}[c]{0.32\textwidth}
		\center
		\includegraphics[width=1\textwidth]{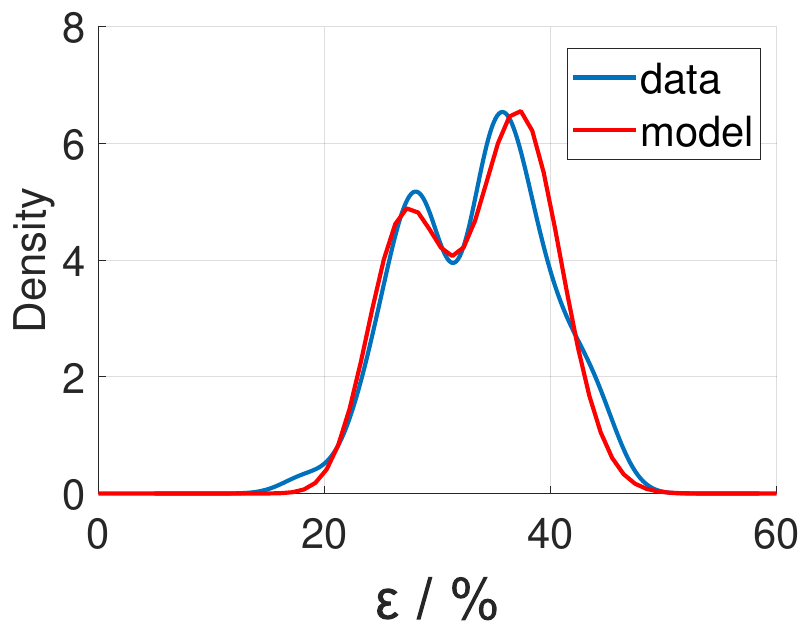}\\
		(b)
	\end{subfigure}
	\begin{subfigure}[c]{0.32\textwidth}
		\center
		\includegraphics[width=1\textwidth]{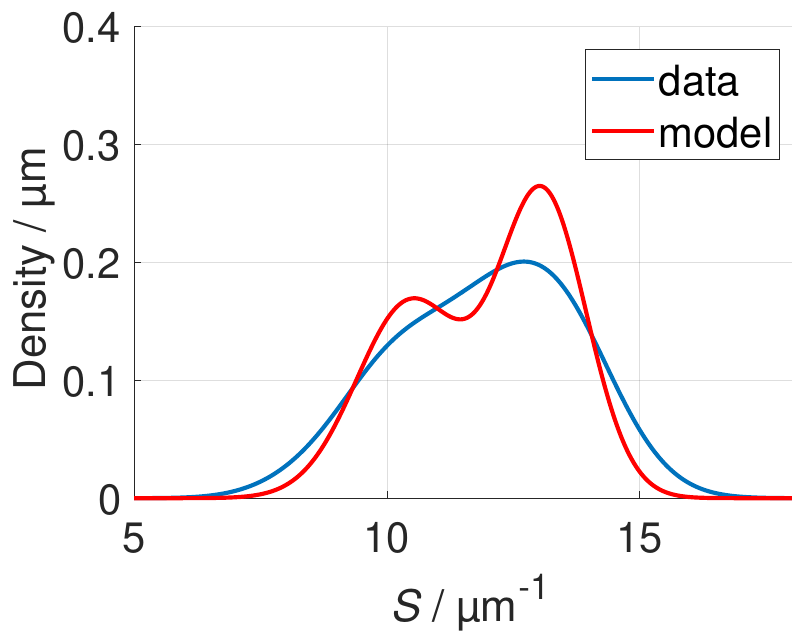}\\
		(c)
	\end{subfigure} 
	\caption{Empirical probability density (data) and fitted parametric density (model) of local thickness $\delta$, local porosity $\varepsilon$, and local surface area per unit volume $S$.}
        \label{fig:univ1}
\end{figure}

We first consider the empirical probability density of the local thickness $\delta$, which is shown in Figure \ref{fig:univ1}a. There is only little variability visible for $\delta$, being almost symmetrically distributed around the mean. As parametric model, a Maxwell-Boltzmann distribution was fitted.
Looking at the results obtained for porosity $\varepsilon$, see Figure~\ref{fig:univ1}b, a bimodal density can be observed. The modes are located at $23\%$ and $35\%$, where the distribution is less concentrated at the latter one. The values of $\varepsilon$ vary from $15\%$ up to $50\%$, which shows a strong heterogeneity of local porosity. Considering the porosity $\varepsilon$ as volume fraction, it takes values between $0$ and $1$. Thus, having in mind that the kernel density estimate of $\varepsilon$ is bimodal, a mixture of two Beta distributions has been chosen as  a parametric model. 
%bimodal?%

A further important morphological descriptor
of battery electrodes is
the surface area of the solid phase, because it heavily influences the electrochemical processes in the electrode. A larger surface area per unit volume, denoted by $S$, can
be advantageous, as it corresponds to the amount of interface at which  electrolyte and  solid
phase can interact. 
The shape of the density of $S$, visible in Figure \ref{fig:univ1}c, results from the correlation between volume fraction and surface area per unit volume, also discussed in the context of the spatial gradients shown in Figure~\ref{fig:globalY}. The surface area per unit volume tends to take its maximum in case that the fraction of solid and pores are nearly identical.
Because of this interdependence of local porosity $\varepsilon$ and local specific surface area $S$ and in view of the bimodality of the density of $\varepsilon$,  a bimodal parametric model, namely a mixture of two Gaussian distributions, was chosen for  $S$. 

%TC:ignore
%TC:endignore

\begin{figure}[ht]
\center
	\begin{subfigure}[c]{0.32\textwidth}
		\center
		\includegraphics[width=1\textwidth]{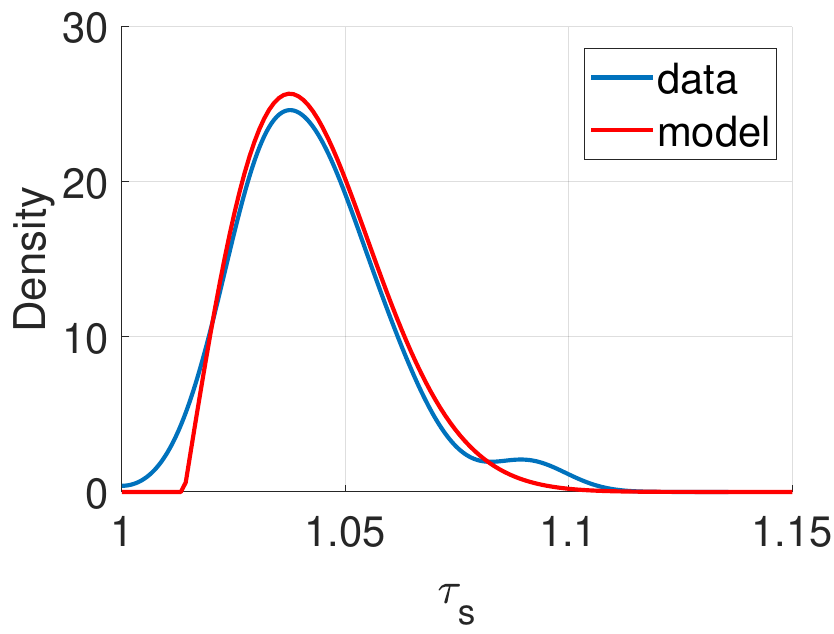}\\
		(a)
	\end{subfigure}
	\begin{subfigure}[c]{0.32\textwidth}
		\center
		\includegraphics[width=1\textwidth]{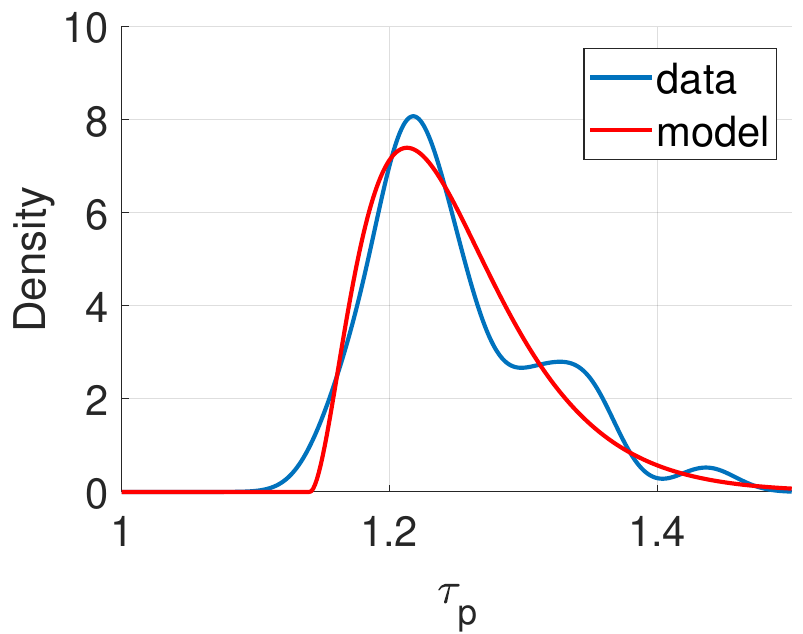}\\
		(b)
	\end{subfigure}
	\caption{Empirical probability density (data) and fitted parametric density (model) of  means geodesic tortuosities, $\tau_s$ and $\tau_p$, in the solid (a) and  pore phase (b), respectively.} \label{fig:univ2}
\end{figure}

Last but not least,  we look at a morphological descriptor which quantifies the lengths of shortest pathways through a given  material phase, namely the so-called mean geodesic tortuosity, see Section~\ref{sec:morphological_descriptors} for details.  
The empirical density of the mean geodesic tortuosity $\tau_s$ in the solid phase is shown in Figure~\ref{fig:univ2}a. For large parts of the electrode, straight pathways in the solid-phase with almost no obstruction from the top- to the bottom-face of the electrode can be found. This results in a distribution, which is mainly concentrated at small tortuosity values of about $1.04$. Note that for PTMA-CMK8 electrodes comparably short--or even shorter--transportation pathways are observed in the solid phase at the microscale~\cite{Ademmer2023}. Compared to nanostructured $\mathrm{Li}\mathrm{Ni}_{1/3}\mathrm{Mn}_{1/3}\mathrm{Co}_{1/3}\mathrm{Co}_{2}$ (NMC) active material particles, 1.04 is rather low. For differently manufactured nanostructured NMC particles, the mean geodesic tortuosity of solid and pores exceeds $1.04$~\cite{neumann_nmc.2023}. Nevertheless, in the present paper, $\tau_s$ still locally reaches values up to $1.1$. This could be explained by the higher porosity in certain areas, which leads to less pathways in the solid phase and, therefore, to larger values of $\tau_s$. Thus, a Rayleigh distribution \cite{siddiqui.1964} was used for the  fitting of a parametric model.
Moreover, the empirical density of mean geodesic tortuosity of  pore phase $\tau_p$, depicted in Figure~\ref{fig:univ2}b, was analyzed. This descriptor is of even higher importance than the mean geodesic tortuosity in the solid phase $\tau_s$, since the supply of electrons in the solid phase is rarely a limiting factor regarding the electrochemical performance of electrodes. The values of $\tau_p$  range from $1.15$ up to $1.5$ with highest concentration at about $1.22$. This effect can again be explained by the local behavior of porosity, since there are in general much fewer possible pathways in the pore space, increasing the shortest path length and, therefore,  increasing the mean geodesic tortuosity $\tau_p$. 
It turned out that a gamma distribution
fits the empirical density of $\tau_p$ quite well, see Figure~\ref{fig:univ2}b. 

%TC:ignore
%TC:endignore

\subsubsection{Bivariate distributions of descriptor pairs}

After having considered the fitting of   univariate probability densities for single morphological descriptors in Section~\ref{sec:univdist}, we   now explain which bivariate parametric
 densities were selected
 for descriptor pairs, consisting of $\varepsilon$ and one of the remaining four descriptors $\delta$, $S$, $\tau_s$ and $\tau_p$. In particular, 
 using bivariate kernel density estimates we fitted one-parametric Archimedean copulas, see  Table~\ref{tab:copula}.

%TC:ignore
\begin{table}[h]
\begin{center}

\begin{tabular}[h]{cc}
2nd descriptor & fitted parameter of Frank copula\\ \hline
$\delta$  &  $\theta= 4.09$ \\ \hline
$S$ & $\theta= 9.77$ \\ \hline
$\tau_s$  & $\theta= 8.89$   \\ \hline
$\tau_p$  & $\theta=  -4.96$ \\

\end{tabular}

\caption{Fitted parameters of the Frank copula for descriptor pairs, where the first descriptor is the  porosity $\varepsilon$.}
\label{tab:copula}
\end{center}
\end{table}
%TC:endignore

\begin{figure}[h]
\center
    \begin{subfigure}[c]{0.24\textwidth}
		\center
		\includegraphics[width=1\textwidth]{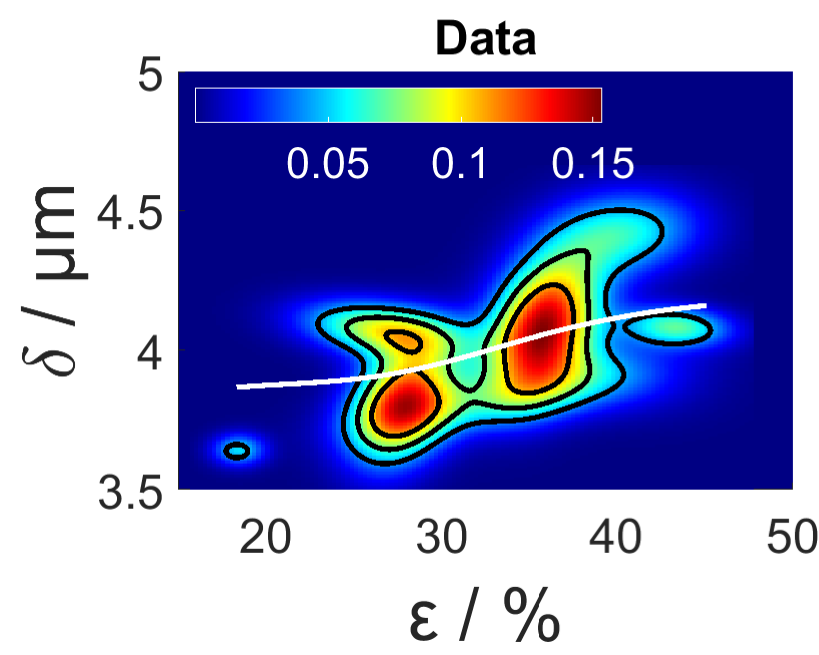}\\
		(a)
	\end{subfigure}
	\begin{subfigure}[c]{0.24\textwidth}
		\center
		\includegraphics[width=1\textwidth]{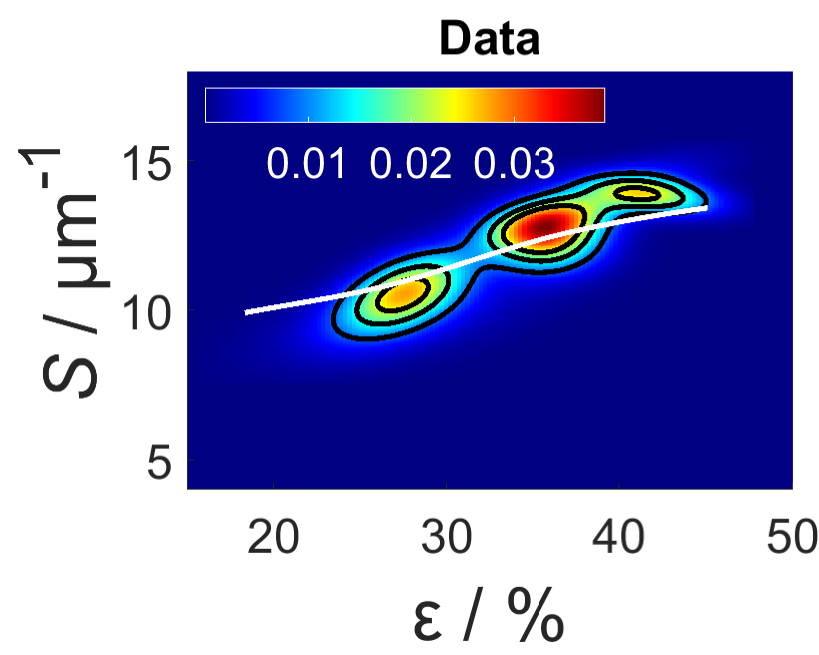}\\
		(b)
	\end{subfigure}
	\begin{subfigure}[c]{0.24\textwidth}
		\center
		\includegraphics[width=1\textwidth]{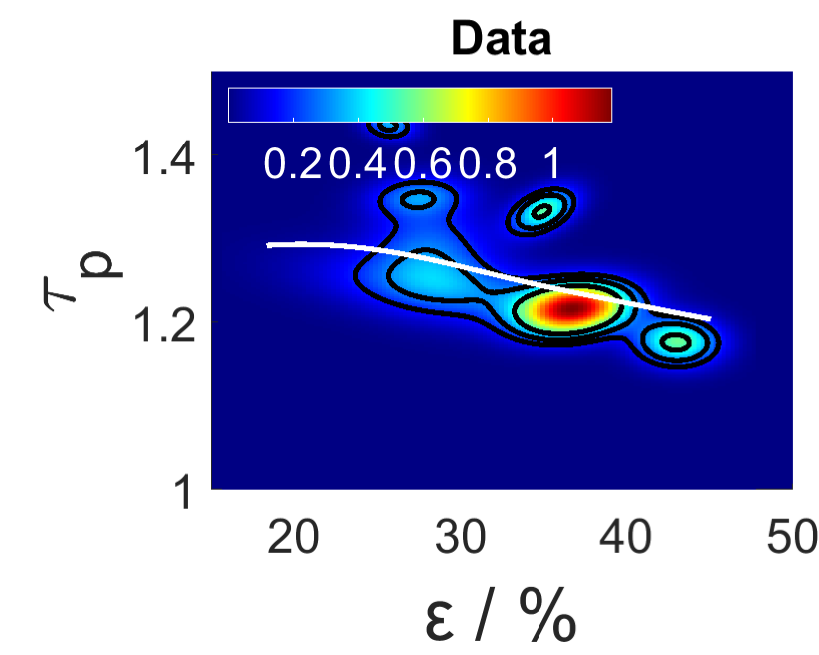}\\
		(c)
	\end{subfigure} 
	\begin{subfigure}[c]{0.24\textwidth}
		\center
		\includegraphics[width=1\textwidth]{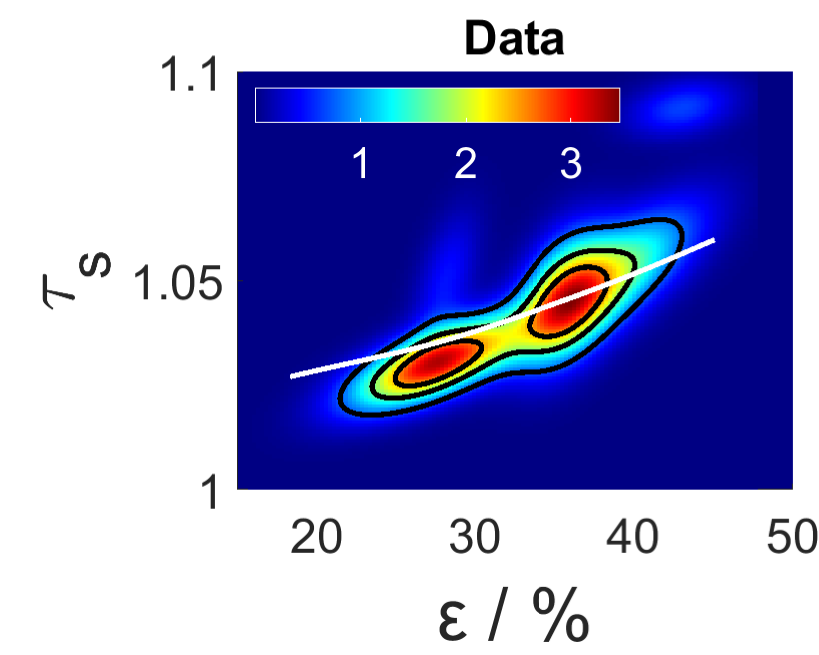}\\
		(d)
	\end{subfigure} 
	
	\begin{subfigure}[c]{0.24\textwidth}
		\center
		\includegraphics[width=1\textwidth]{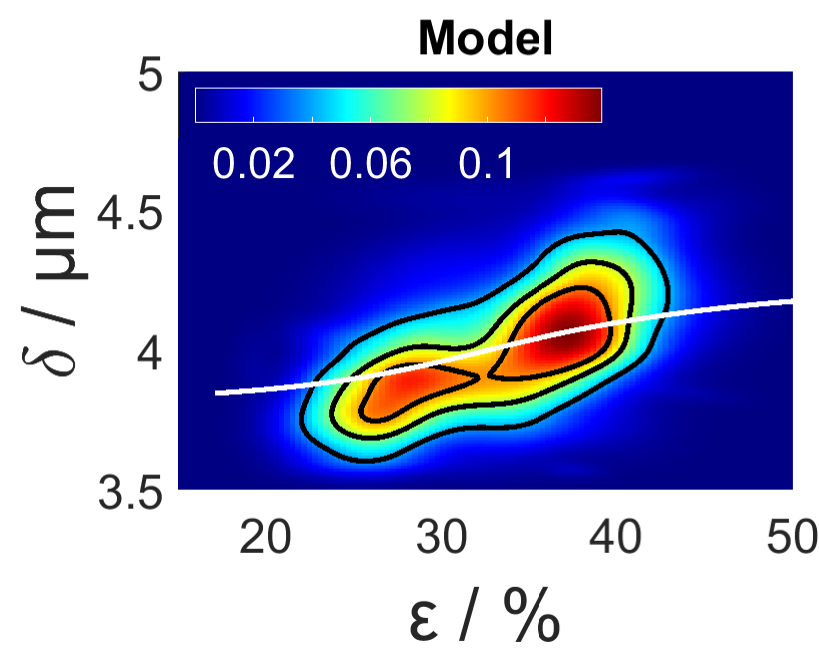}\\
		(e)
	\end{subfigure}
	\begin{subfigure}[c]{0.24\textwidth}
		\center
		\includegraphics[width=1\textwidth]{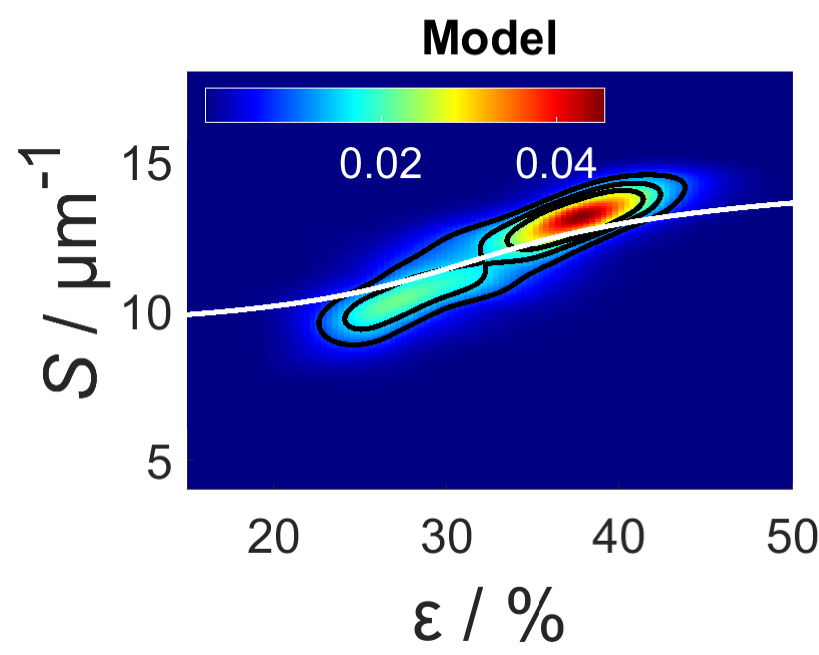}\\
		(f)
	\end{subfigure}
	\begin{subfigure}[c]{0.24\textwidth}
		\center
		\includegraphics[width=1\textwidth]{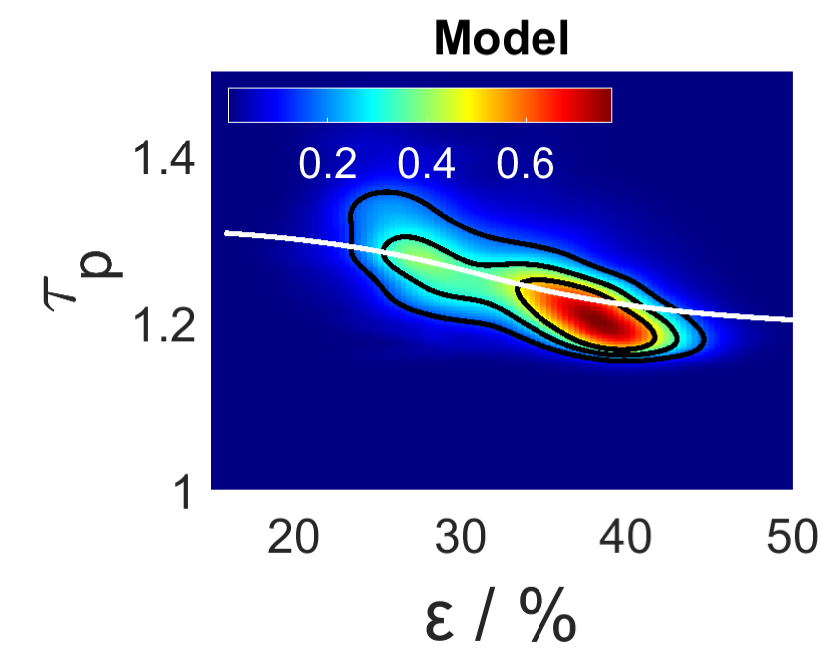}\\
		(g)
	\end{subfigure} 
	\begin{subfigure}[c]{0.24\textwidth}
		\center
		\includegraphics[width=1\textwidth]{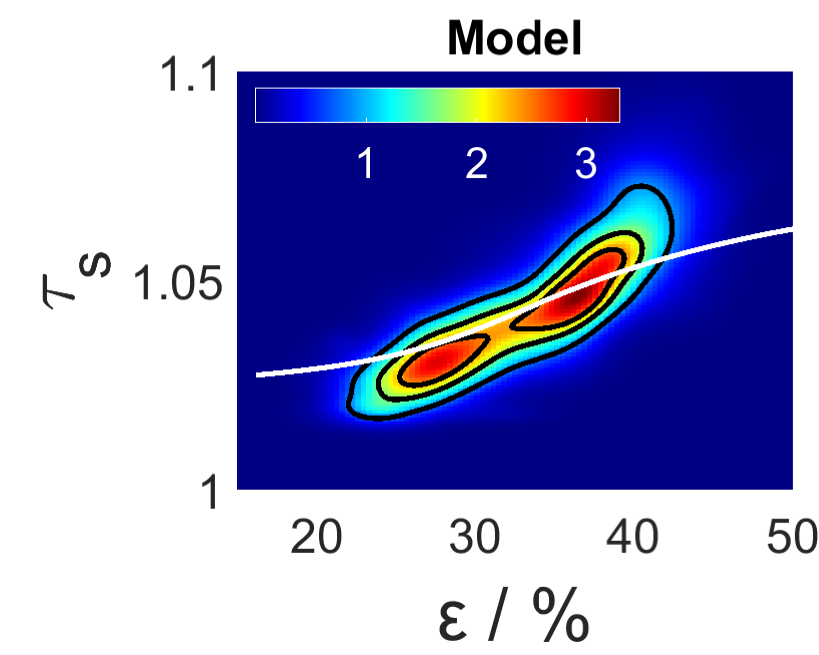}\\
		(h)
	\end{subfigure} 
	\caption{Bivariate empirical  densities (top row) and fitted parametric densities (bottom row) of pairs of local microstructure descriptors. The bivariate probability densities are visualized as heat maps, where the  white lines represent the (conditional) mean values for  given local porosities. The black contour lines are the $25\%$-, $50\%$- and $75\%$-quantiles, respectively.}\label{fig:2DPlots}
\end{figure}

In the next step, using the copula-based representation formula for bivariate probability densities stated in Section~\ref{sec:biv.dis}, two-dimensional density plots for descriptor pairs were determined, where a nice coincidence between empirical  and fitted parametric densities can be observed for all four pairs of local
microstructure descriptors, see Figure~\ref{fig:2DPlots}.  
 Note that black contour lines in the plots resemble the $75\%-,50\%-$ and $25\%-\text{quantiles}$, white lines represent the conditional mean values for  given porosities.

%TC:ignore
%TC:endignore

By visual inspection of the bivariate densities shown in Figure~\ref{fig:2DPlots},
it becomes evident that the local porosity $\varepsilon$ is correlated with the remaining four descriptors
$\delta$, $S$, $\tau_s$ and $\tau_p$.  Regarding the pair $(\varepsilon,\delta)$ of thickness and porosity, considered in Figures~\ref{fig:2DPlots}a and \ref{fig:2DPlots}e, we can see that these two descriptors are positively correlated, see also Table~\ref{tab:correlations-matrix}. This behaviour is sensible in a way that the electrode seems to be thicker in the center of the image data cutout, where we also observed large porous regions compared to other parts of the data, see Figure~\ref{fig:labelling}b.  

Furthermore,  looking at the bivariate density of  $(\varepsilon,S)$, i.e. local porosity and surface area per unit volume, presented  in Figures~\ref{fig:2DPlots}b and \ref{fig:2DPlots}f, we  also observe a positive correlation. 
 On the other hand, the bivariate density of $(\varepsilon,\tau_p)$, i.e. porosity and mean geodesic tortuosity of the pore phase, shown in Figures~\ref{fig:2DPlots}c and \ref{fig:2DPlots}g, indicates
a  negative correlation. This is not surprising, since decreasing the porosity means less pore space and, therefore, fewer pathways that can be traversed, which results in longer pathways and, consequently, increases the mean geodesic tortuosity of the pore space.
  With similar arguments as above, the positive correlation of $\varepsilon$ and $\tau_s$, i.e. local porosity and mean geodesic tortuosity of the solid phase, can be explained. Decreasing porosity increases the volume fraction of solid phase and, thus, the number of possible pathways through the solid phase, which decreases the average length of shortest pathways in this phase.

\subsection{Discussion}\label{sec:dis}
The 3D morphology of the X-PVMPT-based composite electrode investigated in this paper shows two different types of structural heterogeneity. On the one hand, we observed that the porosity of the sample is highly dependent on the $z$-position in the electrode, meaning that the electrode material is more porous near the top face. These porous regions at the top also have coarser surfaces, since the boundary area of the solid phase is larger at the top face of the electrode, see Figure~\ref{fig:globalY}. These spatial gradients might presumably arise during the drying process. In order to support this conjecture and to achieve a detailed understanding of the relationships between process parameters and the gradient of porosity, further research is required including 3D imaging of differently manufactured X-PVMPT electrodes.

On the other hand, we observed considerable local heterogeneities of the electrode material for different 3D cutouts along the $(x,y)$-plane. This can be seen by the high variances of the univariate distributions of the local morphological descriptors fitted  in Section~\ref{sec:univdist} to segmented image data.
Furthermore, we observed a positive correlation between local  porosity $\varepsilon$ and  thickness $\delta$ as well as between $\varepsilon$ and surface area per unit volume $S$.  
 However, note that the values of local thickness $\delta$ of the electrode material show relatively little variation and  are pretty closely  grouped around their mean value, see Figure~\ref{fig:univ1}a.
The mean geodesic tortuosity of the solid phase $\tau_s$ showed values close to 1, which implicates short transportation paths trough the solid phase. 
In contrast we have seen that the mean geodesic tortuosity of the pore phase $\tau_p$ exhibits larger values, which tells us that the paths through the porous phase are much more winded.

\section{Conclusion}\label{sec:conclusion}

Our analysis of the X-PVMPT-based composite electrodes provides valuable insights into the nanostructural properties and local heterogeneities of its 3D morphology. 
The combination of FIB-SEM tomography, which achieved a voxel resolution of 10 nm, and image-segmentation techniques allowed for a detailed morphological characterization of the electrode. The investigations performed in this study focused on key morphological descriptors such as thickness, porosity, surface area per unit volume, and mean geodesic tortuosity for both, solid and pore phases. These descriptors are crucial for the performance of the electrode. In particular, the porosity was observed to be highly variable and dependent on the $z$-position within the electrode, with higher porosity and coarser surfaces towards the top. This heterogeneity was further supported by the distributions of local descriptors, which were modeled using parametric probability distributions. The bivariate densities, modeled with Frank copulas, revealed quantitative relationships between descriptors, especially the correlations between porosity and the remaining four morphological descriptors. 
Notably, the values obtained for the local mean geodesic tortuosities $\tau_s$ and $\tau_p$ indicated that pathways through the solid phase were relatively straight, whereas those through the pore phase were more winded, impacting the overall charge-transport properties through these two phases. 
These insights emphasize the critical role of the nanostructure with respect to the performance of organic electrode-active materials (OAMs). Understanding and controlling these morphological features can pave the way for optimizing battery design for X-PVMPT electrodes and enhancing the efficiency of polymer-based batteries, as it was already done for NMC-based electrodes, see, e.g.~\cite{kremer.2020}. The results obtained in the present paper provide a foundation for future research on the morphological optimization of OAM-based electrodes, contributing to the advancement of safer, more efficient, and cost-effective battery technologies.

%TC:ignore
\section*{Acknowledgement}

This research has been  supported by the German Research Foundation (DFG) under grants 441292784 and 
441215516, within Priority program
2248 ``Polymer-based batteries'' of DFG. 
Moreover, this research was funded by DFG under Project ID 390874152 (POLiS Cluster of Excellence, EXC 2154). The present paper contributes to the research performed at CELEST (Center for Electrochemical Energy Storage Ulm-Karlsruhe).

\newpage

\setcounter{page}{1}
\renewcommand\thepage{S\arabic{page}}
\section*{Supplementary information}

\subsection*{Details of image segmentation}

For image segmentation, i.e., for classifying the voxels as pores or solid, the following features are used: Gaussian smoothing, Laplacian of Gaussian, Gaussian gradient magnitude, difference of Gaussians, structure tensor eigenvalues, and Hessian of Gaussian eigenvalues, each one for $\sigma \in \{0.3,0.7,1.0,1.6,3.5,5.0,10.0\}$. The output of the trained random forests is a probability map, indicating the probability whether a voxel belongs to the pore space or to the solid phase. Each voxel that belongs to the solid phase with probability greater than 0.4 is assigned to solid phase and otherwise to background, leads to a segmentation into the solid phase and its complement. Note that this threshold has been manually determined
based on visual inspection, where we used the open-source software Fiji. 

\subsection*{Univariate probability density functions}

The probability density functions $f:\R\to[0,\infty)$ of the distributions introduced in Section~\ref{sec:uni.den.sin} and used to model the univariate distributions of local morphological descriptors in Section~\ref{sec:univdist} are provided in the following.

\begin{enumerate}

\item Mixture of two Beta distributions $\mathsf{Beta}(\alpha_1, \alpha_2, \beta_1, \beta_2, p_B)$ 
 \begin{equation*}
    f\left(x\right) = p_B\;\frac{\Gamma\left(\alpha_1+\beta_1\right)x^{\alpha_1-1}\left(1-x\right)^{\beta_1-1}}{\Gamma\left(\alpha_1\right)\Gamma\left(\beta_1\right)} +\left(1-p_B\right)\;\frac{\Gamma\left(\alpha_2+\beta_2\right)x^{\alpha_2-1}\left(1-x\right)^{\beta_2-1}}{\Gamma\left(\alpha_2\right)\Gamma\left(\beta_2\right)}, 
\end{equation*} 
for each $x \in [0,1]$, where $\alpha_1, \alpha_2, \beta_1, \beta_2 > 0$ and $0 \leq p_B \leq 1$.

\item
Mixture of two Gaussian distributions $\mathsf{N}(\mu_1, \mu_2, \sigma_1, \sigma_2, p_N)$

\begin{equation*}
    f\left(x\right) = p_N\;\frac{1}{\sigma_1\sqrt{2 \pi}}\exp\left(\frac{\left(x-\mu_1\right)^2}{2\sigma_1^2}\right) + \left(1-p_N\right)\;\frac{1}{\sigma_2\sqrt{2\pi}}\exp\left(\frac{\left(x-\mu_2\right)^2}{2\sigma_2^2}\right)
\end{equation*}
for each $x \in \R$, where $\mu_1, \mu_2 \in \R, \sigma_1, \sigma_2 > 0$ and $0 < p_N < 1.$

\item
Maxwell-Boltzmann distribution $\mathsf{MB}(\mu_{M}, \sigma_{M})$
\begin{equation*}
    f\left(x\right) = \sqrt{\frac{2}{\pi}}\frac{(x-\mu_{M})^2}{\sigma_{M}^2}\exp\left(-\frac{(x-\mu_{M})^2}{2\sigma_{M}^2}\right), 
\end{equation*} 
for each $x>\mu_{M}$, where $\mu_{M} \in \R$ and $\sigma_{M} > 0.$

\item
Shifted Gamma distribution $\Gamma(a_\Gamma, \sigma_\Gamma, \mu_\Gamma)$
\begin{equation*}
     f\left(x\right) = \frac{1}{\Gamma\left(a_\Gamma\right)}\left(\frac{x-\mu_\Gamma}{\sigma_\Gamma}\right)^{a_\Gamma-1}\exp\left(-\frac{x-\mu_\Gamma}{\sigma_\Gamma}\right),
 \end{equation*}
for each $x>\mu_\Gamma,$ where $a_\Gamma, \sigma_\Gamma > 0$ and $\mu_\Gamma \in \R.$

\item
Rayleigh distribution $\mathsf{R}(\mu_R, \sigma_R)$
\begin{equation*}
    f\left(x\right) = \frac{x-\mu_R}{\sigma_R}\exp\left(-\frac{(x-\mu_R)^2}{2\sigma_R^2}\right),
\end{equation*}
for each $x > \mu_R,$ where $\mu_R \in \R$ and $\sigma_R > 0.$

\subsection*{Brief introduction to copulas}

We briefly recall the concept of copulas.  Let $\left(U,V\right)$ be a two-dimeensional random vector taking values in the unit square $[0,1]^2$, where $U$ and $V$ are uniformly distributed on the unit interval $[0,1]$. Then, the joint probability distribution function $C: [0,1]^2 \rightarrow [0,1]$  of $\left(U,V\right)$ is called a two-dimensional copula, where $C(u,v) = P(U \leq u,V \leq v)$ with $P(U\leq u) = u$ and $P(V \leq v)= v$ for any $u,v\in [0,1]$.  
From Sklar's representation formula~\cite{nelsen.2007} we get that for any two-dimensional random vector $(X,Y)$, its  joint probability distribution function $H:\R^2\to[0,1]$ with $H(x,y)=P(X\le x, Y\le y)$  can be written in the form
$$
H(x,y) = C\left(F\left(x\right),G\left(y\right)\right)
$$
for all $x,y \in \mathbb{R}$, where $F:\mathbb{R} \rightarrow [0,1]$ with $F\left(x\right) = P\left(X\leq x\right)$ for each $x \in \R$ and $G:\mathbb{R} \rightarrow [0,1]$ with $G\left(y\right) = P\left(Y\leq y\right)$ for each $y \in \R$ are the univariate distribution functions of $X$ and $Y$, respectively, and $C:[0,1]^2\to[0,1]$ is a certain copula. Moreover, if the functions $F,G$ and $C$ are differentiable, then    the joint probability density $h:\mathbb{R}^2 \rightarrow [0,\infty)$  of $(X,Y)$  can be written as
\begin{equation*}\label{eq:bivariate}
    h(x,y) = f(x)g(y)\left(\frac{\partial^2}{\partial x\partial y}C\right)(F(x),G(y)) 
\end{equation*}
for all $x,y \in \mathbb{R}$, where $f:\mathbb{R} \rightarrow [0,\infty)$ and $g:\mathbb{R} \rightarrow [0,\infty)$ are the univariate probability densities of $X$ and $Y$, respectively. From the joint probability density, we directly obtain the conditional probability density $h_{Y=y}:\mathbb{R} \rightarrow [0, \infty)$ of $X$ given $Y=y$ for each $y \in \R$ fulfilling $g(y) > 0$. It reads as
$$h_{Y=y}(x) = f(x)\left(\frac{\partial^2}{\partial x\partial y}C\right)(F(x),G(y))$$ for each $ x \in \mathbb{R}.$

\end{enumerate}

%\newpage

%\input{Reply}

%TC:endignore

\begin{thebibliography}{10}

\bibitem{kimJ1}
J.~Kim, Y.~Kim, G.~Kwon, Y.~Ko, and K.~Kang.
\newblock Organic batteries for a greener rechargeable world.
\newblock {\em Nature Reviews Materials}, 8:54–70, 2022.

\bibitem{esser2}
B.~Esser.
\newblock Redox polymers as electrode-active materials for batteries.
\newblock {\em Organic Materials}, 1:063--070, 2019.

\bibitem{esser3}
B.~Esser, F.~Dolhem, M.~Becuwe, P.~Poizot, A.~Vlad, and D.~Brandell.
\newblock A perspective on organic electrode materials and technologies for next generation batteries.
\newblock {\em Power Sources}, 482, 2021.

\bibitem{Poizot4}
P.~Poizot, J.~Gaubicher, S.~Renault, L.~Dubois, Y.~Liang, and Y.~Yao.
\newblock Opportunities and challenges for organic electrodes in electrochemical energy storage.
\newblock {\em Chemical Reviews}, 120(14):6490--6557, 2020.

\bibitem{yang5}
H.~Yang, J.~Lee, J.~Y. Cheong, Y.~Wang, G.~Duan, H.~Hou, S.~Jiang, and I.-D. Kim.
\newblock Molecular engineering of carbonyl organic electrodes for rechargeable metal-ion batteries: fundamentals, recent advances, and challenges.
\newblock {\em Energy \& Environmental Science}, 14:4228--4267, 2021.

\bibitem{huang6}
Z.~Huang, X.~Du, M.~Ma, S.~Wang, Y.~Xie, Y.~Meng, W.~You, and L.~Xiong.
\newblock Organic cathode materials for rechargeable aluminum-ion batteries.
\newblock {\em ChemSusChem}, 16(9):637--671, 2023.

\bibitem{li7}
Z.~Li, J.~Häcker, M.~Fichtner, and Z.~Zhao-Karger.
\newblock Cathode materials and chemistries for magnesium batteries: challenges and opportunities.
\newblock {\em Advanced Energy Materials}, 13(27), 2023.

\bibitem{chen8}
Y.~Chen, K.~Fan, Y.~Gao, and C.~Wang.
\newblock Challenges and perspectives of organic multivalent metal-ion batteries.
\newblock {\em Advanced Materials}, 34(52):2200662, 2022.

\bibitem{poizot9}
P.~Poizot, F.~Dolhem, and J.~Gaubicher.
\newblock Progress in all-organic rechargeable batteries using cationic and anionic configurations: toward low-cost and greener storage solutions?
\newblock {\em Current Opinion in Electrochemistry}, 9:70--80, 2018.

\bibitem{Kolek10}
M.~Kolek, F.~Otteny, P.~Schmidt, C.~Mück-Lichtenfeld, C.~Einholz, J.~Becking, E.~Schleicher, M.~Winter, P.~Bieker, and B.~Esser.
\newblock Ultra-high cycling stability of poly(vinylphenothiazine) as a battery cathode material resulting from $\pi$-$\pi$ interactions.
\newblock {\em Energy \& Environmental Science}, 10:2334--2341, 2017.

\bibitem{kolek11}
M.~Kolek, F.~Otteny, J.~Becking, M.~Winter, B.~Esser, and P.~Bieker.
\newblock Mechanism of charge/discharge of poly(vinylphenothiazine)-based li–organic batteries.
\newblock {\em Chemistry of Materials}, 30(18):6307--6317, 2018.

\bibitem{otteny12}
F.~Otteny, V.~Perner, C.~Einholz, G.~Desmaizieres, E.~Schleicher, M.~Kolek, P.~Bieker, M.~Winter, and B.~Esser.
\newblock Bridging the gap between small molecular $\pi$-interactions and their effect on phenothiazine-based redox polymers in organic batteries.
\newblock {\em ACS Applied Energy Materials}, 4(8):7622--7631, 2021.

\bibitem{perner13}
V.~Perner, D.~Diddens, F.~Otteny, V.~Küpers, P.~Bieker, B.~Esser, M.~Winter, and M.~Kolek.
\newblock Insights into the solubility of poly(vinylphenothiazine) in carbonate-based battery electrolytes.
\newblock {\em ACS Applied Materials \& Interfaces}, 13(10):12442--12453, 2021.

\bibitem{tengen14}
B.~Tengen, T.~Winkelmann, N.~Ortlieb, V.~Perner, G.~Studer, M.~Winter, B.~Esser, A.~Fischer, and P.~Bieker.
\newblock Immobilizing poly(vinylphenothiazine) in ketjenblack-based electrodes to access its full specific capacity as battery electrode material.
\newblock {\em Advanced Functional Materials}, 33(9):2210512, 2023.

\bibitem{otteny15}
F.~Otteny, M.~Kolek, J.~Becking, M.~Winter, P.~Bieker, and B.~Esser.
\newblock Unlocking full discharge capacities of poly(vinylphenothiazine) as battery cathode material by decreasing polymer mobility through cross-linking.
\newblock {\em Advanced Energy Materials}, 8(33):1802151, 2018.

\bibitem{studer16}
G.~Studer, A.~Schmidt, J.~Büttner, M.~Schmidt, A.~Fischer, I.~Krossing, and B.~Esser.
\newblock On a high-capacity aluminium battery with a two-electron phenothiazine redox polymer as a positive electrode.
\newblock {\em Energy \& Environmental Science}, 16:3760--3769, 2023.

\bibitem{bhosale17}
M.~Bhosale, C.~Schmidt, P.~Penert, G.~Studer, and B.~Esser.
\newblock Anion-rocking chair batteries with tuneable voltage using viologen- and phenothiazine polymer-based electrodes**.
\newblock {\em ChemSusChem}, 17(5):e202301143, 2024.

\bibitem{desmaizieres18}
G.~Desmaizieres, V.~Perner, D.~Wassy, M.~Kolek, P.~Bieker, M.~Winter, and B.~Esser.
\newblock Evaluating the polymer backbone – vinylene versus styrene – of anisyl-substituted phenothiazines as battery electrode materials.
\newblock {\em Batteries \& Supercaps}, 6(2):e202200464, 2023.

\bibitem{otteny19}
F.~Otteny, G.~Studer, M.~Kolek, P.~Bieker, M.~Winter, and B.~Esser.
\newblock Phenothiazine-functionalized poly(norbornene)s as high-rate cathode materials for organic batteries.
\newblock {\em ChemSusChem}, 13(9):2232--2238, 2020.

\bibitem{acker20}
P.~Acker, L.~Rzesny, C.r F.~N. Marchiori, C.~M. Araujo, and B.~Esser.
\newblock $\pi$-conjugation enables ultra-high rate capabilities and cycling stabilities in phenothiazine copolymers as cathode-active battery materials.
\newblock {\em Advanced Functional Materials}, 29(45):1906436, 2019.

\bibitem{acker21}
P.~Acker, J.~S. Wössner, G.~Desmaizieres, and B.~Esser.
\newblock Conjugated copolymer design in phenothiazine-based battery materials enables high mass loading electrodes.
\newblock {\em ACS Sustainable Chemistry \& Engineering}, 10(10):3236--3244, 2022.

\bibitem{wessling22}
R.~Wessling, R.~Delgado~Andrés, I.~Morhenn, P.~Acker, W.~Maftuhin, M.~Walter, U.~Würfel, and B.~Esser.
\newblock Phenothiazine-based donor–acceptor polymers as multifunctional materials for charge storage and solar energy conversion.
\newblock {\em Macromolecular Rapid Communications}, 45(1):2200699, 2024.

\bibitem{Delgado23}
R.~Delgado~Andrés, R.~Wessling, J.~Büttner, L.~Pap, A.~Fischer, B.~Esser, and U.~Würfel.
\newblock Organic photo-battery with high operating voltage using a multi-junction organic solar cell and an organic redox-polymer-based battery.
\newblock {\em Energy \& Environmental Science}, 16:5255--5264, 2023.

\bibitem{wessling.2024}
R.~Wessling, H.~Koger, F.~Otteny, M.~Schmidt, A.~Semmelmaier, and B.~Esser.
\newblock Unlocking twofold oxidation in phenothiazine polymers for application in symmetric all-organic anionic batteries.
\newblock {\em ACS Applied Polymer Materials}, 6:7956--7968, 2024.

\bibitem{otteny24}
F.~Otteny, G.~Desmaizieres, and B.~Esser.
\newblock {Phenothiazine-based redox polymers for energy storage}.
\newblock In {\em {Redox Polymers for Energy and Nanomedicine}}, pages 166 -- 197. The Royal Society of Chemistry, 2020.

\bibitem{Neumann23}
M.~Neumann, M.~Ademmer, M.~Osenberg, A.~Hilger, F.~Wilde, S.~Muench, M.D. Hager, U.S. Schubert, I.~Manke, and V.~Schmidt.
\newblock {3D} microstructure characterization of polymer battery electrodes by statistical image analysis based on synchrotron x-ray tomography.
\newblock {\em Journal of Power Sources}, 542, 2022.

\bibitem{Ademmer2023}
M.~Ademmer, P.-H. Su, L.~Dodell, J.~Asenbauer, M.~Osenberg, A.~Hilger, I.~Chang, J.-K.and~Manke, M.~Neumann, V.~Schmidt, and D.~Bresser.
\newblock Unveiling the impact of crosslinking redox-active polymers on their electrochemical behavior by {3D} imaging and statistical microstructure analysis.
\newblock {\em The Journal of Physical Chemistry C}, 127:19366--19377, 2023.

\bibitem{heenan.2019}
T.~M.~M. Heenan, C.~Tan, J.~Hack, D.~J.~L. Brett, and R.~P. Shearing.
\newblock Developments in {X-ray} tomography characterization for electrochemical devices.
\newblock {\em Materials Today}, 31:69--85, 2019.

\bibitem{tang.2021}
F.~Tang, Z.~Wu, C.~Yang, M.~Osenberg, A.~Hilger, K.~Dong, H.~Markötter, I.~Manke, F.~Sun, L.~Chen, and G.~Cui.
\newblock Synchrotron {X-Ray} tomography for rechargeable battery research: fundamentals, setups and applications.
\newblock {\em Small Methods}, 5(9):2100557, 2021.

\bibitem{holzer.2012}
L.~Holzer and M.~Cantoni.
\newblock Review of {FIB}-tomography.
\newblock In I.~Utke, S.~Moshkalev, and P.~Russell, editors, {\em Nanofabrication using Focused Ion and Electron Beams: Principles and Applications}, pages 410--435. Oxford University Press, 2012.

\bibitem{moebus.2007}
G.~M{\"o}bus and B.~J. Inkson.
\newblock Nanoscale tomography in materials science.
\newblock {\em Materials Today}, 10:18--25, 2007.

\bibitem{berg.2019}
S.~Berg, D.~Kutra, T.~Kroeger, C.~N. Straehle, B.~X. Kausler, C.~Haubold, M.~Schiegg, J.~Ales, T.~Beier, M.~Rudy, K.~Eren, J.~I. Cervantes, B.~Xu, F.~Beuttenmueller, A.~Wolny, C.~Zhang, U.~Koethe, F.~A. Hamprecht, and A.~Kreshuk.
\newblock ilastik: interactive machine learning for (bio)image analysis.
\newblock {\em Nature Methods}, 16:1226—1232, 2019.

\bibitem{n.2020}
M.~Neumann, E.~Machado~Charry, K.~Zojer, and V.~Schmidt.
\newblock On variability and interdependence of local porosity and local tortuosity in porous materials: a case study for sack paper.
\newblock {\em Methodology and Computing in Applied Probability}, 23:613–627, 2021.

\bibitem{neumann.2024}
M.~Neumann, P.~Gr{\"a}fensteiner, E.~Machado~Charry, U.~Hirn, A.~Hilger, I.~Manke, R.~Schennach, V.~Schmidt, and K.~Zojer.
\newblock R-vine copulas for data-driven quantification of descriptor relationships in porous materials.
\newblock {\em Advanced Theory and Simulations}, 2024:2301261, 2024.

\bibitem{lowe.2004}
D.~G. Lowe.
\newblock Distinctive image features from scale-invariant keypoints.
\newblock {\em International Journal of Computer Vision}, 60:91--110, 2004.

\bibitem{Fiji2012}
J.~Schindelin, I.~Arganda-Carreras, E.~Frise, V.~Kaynig, M.~Longair, T.~Pietzsch, S.~Preibisch, C.~Rueden, S.~Saalfeld, B.~Schmid, J.-Y. Tinevez, D.~J. White, V.~Hartenstein, K.~Eliceiri, P.~Tomancak, and A.~Cardona.
\newblock Fiji: an open-source platform for biological-image analysis.
\newblock {\em Nature Methods}, 9:676–682, 2012.

\bibitem{breiman.2001}
L.~Breiman.
\newblock Random forests.
\newblock {\em Machine Learning}, 45:5--32, 2001.

\bibitem{machado}
E.~Machado~Charry, M.~Neumann, J.~Lahti, R.~Schennach, V.~Schmidt, and K.~Zojer.
\newblock Pore space extraction and characterization of sack paper using \textmu-\text{CT}.
\newblock {\em Journal of Microscopy}, 272:35--46, 2018.

\bibitem{holzer.2023}
L.~Holzer, P.~Marmet, M.~Fingerle, A.~Wiegmann, M.~Neumann, and V.~Schmidt.
\newblock {\em Tortuosity and Microstructure Effects in Porous Media: Classical Theories, Empirical Data and Modern Methods}.
\newblock Springer, 2023.

\bibitem{chiu2013stochastic}
S.N. Chiu, D.~Stoyan, W.S. Kendall, and J.~Mecke.
\newblock {\em Stochastic Geometry and Its Applications}.
\newblock J. Wiley \& Sons, 3rd edition, 2013.

\bibitem{schladitz}
J.~Ohser and K.~Schladitz.
\newblock {\em {3D} Images of Materials Structures: Processing and Analysis}.
\newblock Wiley-{VCH}, 2009.

\bibitem{clennell.1997}
M.~B. Clennell.
\newblock Tortuosity: a guide through the maze.
\newblock {\em Geological Society, London, Special Publications}, 122:299--344, 1997.

\bibitem{ghanbarian.2013}
B.~Ghanbarian, A.~G. Hunt, R.~P. Ewing, and M.~Sahimi.
\newblock Tortuosity in porous media: a critical review.
\newblock {\em Soil Science Society of America Journal}, 77(5):1461--1477, 2013.

\bibitem{neumann.2020}
M.~Neumann, O.~Stenzel, F.~Willot, L.~Holzer, and V.~Schmidt.
\newblock Quantifying the influence of microstructure on effective conductivity and permeability: virtual materials testing.
\newblock {\em International Journal of Solid and Structures}, 184:211--220, 2020.

\bibitem{prifling.2021b}
B.~Prifling, M.~Röding, P.~Townsend, M.~Neumann, and V.~Schmidt.
\newblock Large-scale statistical learning for mass transport prediction in porous materials using 90,000 artificially generated microstructures.
\newblock {\em Frontiers in Materials}, 8:786502, 2021.

\bibitem{thulasiraman1992graphs}
K.~Thulasiraman and M.N.S. Swamy.
\newblock {\em Graphs: Theory and Algorithms}.
\newblock J. Wiley \& Sons, 1st edition, 1992.

\bibitem{hamparsum.1987}
H.~Bozdogan.
\newblock Model selection and {A}kaike's information criterion ({AIC}): The general theory and its analytical extensions.
\newblock {\em Psychometrika}, 52:345--370, 1987.

\bibitem{nelsen.2007}
R.~B. Nelsen.
\newblock {\em An {Introduction} to {Copulas}}.
\newblock Springer, 2007.

\bibitem{neumann_nmc.2023}
M.~Neumann, S.~E. Wetterauer, M.~Osenberg, A.~Hilger, P.~Gräfensteiner, A.~Wagner, N.~Bohn, J.~R. Binder, I.~Manke, T.~Carraro, and V.~Schmidt.
\newblock A data-driven modeling approach to quantify morphology effects on transport properties in nanostructured {NMC} particles.
\newblock {\em International Journal of Solids and Structures}, 280:112394, 2023.

\bibitem{siddiqui.1964}
M.~M. Siddiqui.
\newblock Statistical inference for rayleigh distributions.
\newblock {\em Journal of Research of the National Bureau of Standards, Sec. D}, 68(9):1005--1010, 1964.

\bibitem{kremer.2020}
L.~S. Kremer, A.~Hoffmann, T.~Danner, S.~Hein, B.~Prifling, D.~Westhoff, C.~Dreer, A.~Latz, V.~Schmidt, and M.~Wohlfahrt-Mehrens.
\newblock Manufacturing process for improved ultra-thick cathodes in high-energy lithium-ion batteries.
\newblock {\em Energy Technology}, 8(2):1900167, 2020.

\end{thebibliography}
\end{document}